\documentclass{emulateapj}

\usepackage{graphicx}
\newcommand{\be}{\begin{equation}}
\newcommand{\ee}{\end{equation}}
\newcommand{\bea}{\begin{eqnarray}}
\newcommand{\eea}{\end{eqnarray}}

\begin{document}
\title{Galactic foregrounds: 
Spatial fluctuations and a procedure of removal}
\shorttitle{Synchrotron and Dust Foregrounds}
\shortauthors{Cho \& Lazarian}

\author{Jungyeon Cho\altaffilmark{1,2} and
A. Lazarian\altaffilmark{2}
}
\altaffiltext{1}{Dept. of Astronomy and Space Science, 
    Chungnam National Univ., Daejeon, Korea; cho@canopus.cnu.ac.kr}
\altaffiltext{2}{Dept. of Astronomy, Univ. of Wisconsin, Madison, 
    WI53706, USA; cho@astro.wisc.edu, lazarian@astro.wisc.edu}

\begin{abstract}
Present day cosmic microwave background (CMB) studies require more accurate  
removal of Galactic foreground emission. 
This removal becomes even more essential for CMB polarization measurements.
In this paper, we consider a way of filtering out the diffuse Galactic
fluctuations on the basis of their statistical properties, namely, the power-law spectra of fluctuations.
We focus on the statistical properties of two major Galactic foregrounds that arise from
magnetized turbulence, namely, diffuse synchrotron emission and
thermal emission from dust and describe how their power laws change with the
Galactic latitude. 
We attribute this change to the change of the geometry of the emission region
and claim that the universality of the turbulence spectrum provides a new way of removing Galactic foregrounds.
For the Galactic synchrotron emission, we mainly focus on the geometry of the synchrotron emitting regions, 
which will provide useful information for future polarized synchrotron emission studies. 
Our model calculation suggests that either a one-component extended halo model or a two-component model, 
an extended halo component (scale height $\gtrsim 1kpc$) plus 
a local component, can explain the observed angular spectrum of
the synchrotron emission.
For thermal emission from Galactic dust, we discuss general properties of
a publicly available
94GHz total dust emission map and explain how we can obtain
a polarized dust emission map. Based on a simple model calculation, 
we obtain the angular spectrum of the polarized dust emission.
Our model calculation suggests that $C_l\propto l^{-11/3}$
for $l\gtrsim 1000$ and a shallower spectrum for $l\lesssim 1000$.
We discuss and demonstrate how we can make use of our findings to remove
Galactic foregrounds using a template of spatial fluctuations. In particular, 
we consider examples of spatial filtering of a foreground at small scales, when the 
separation into CMB signal and foregrounds is done at larger scales. 
We demonstrate that the new technique of spatial filtering of foregrounds may be
promising for recovering the CMB signal in a situation when foregrounds are
known at a scale different from the one under study. 
It can also improve filtering by combining measurements obtained at different scales.
\end{abstract}

\section{Introduction}

An important problem in the studies of the early universe with the 
CMB fluctuations is related to separating them from Galactic foregrounds. 
The techniques of removing foregrounds are rather elaborate, 
but in most cases they include using the frequency templates of foreground emission. 
This requires multi-frequency measurements, which are not always available. 
Moreover, some foregrounds, e.g. the so-called spinning dust 
(Draine \& Lazarian 1998ab; Finkbeiner et al. 2002; Lazarian \& Finkbeiner  2003), 
demonstrate rather complex frequency dependence. 
Due to the utmost importance of obtaining CMB signal free of contamination, 
it is essential to consider other ways
to remove foregrounds. 
One way to do this is to take into account the known spatial properties of emission. 

If a foreground has well-defined statistics of spatial fluctuations, 
one can devise techniques of removing the contribution of the foreground 
to the measured microwave signal (see \S 2). 
The issue in this case is whether foregrounds have well defined behavior 
in terms of their spatial statistics. 
Determining this with the available data is the first thrust of our present study 
which we pursue using the Galactic synchrotron and the Galactic dust emission data. 
The second thrust is devising possible ways of removing the foregrounds using 
the self-similarity of Galactic turbulence that gives rise to the foreground fluctuations. 

This work continues our brief study in Cho \& Lazarian (2002a, henceforth CL02) 
where we argued that the properties of Galactic foreground radiation can be 
explained on accepting that the interstellar medium (ISM) that provides 
the fluctuations 
is turbulent and therefore the spatial fluctuations of foregrounds inherit 
power-law spectra of the underlying magnetohydrodynamic (MHD) turbulence. 
Since that work,  better understanding of the properties of the MHD turbulence 
has been achieved. 
For instance, it became clear that the spectrum of density in compressible 
MHD turbulence can be substantially shallower than the Kolmogorov spectrum 
that we assumed in CL02 (see Beresnyak, Lazarian \& Cho 2005). 
Moreover, the search for alternative procedures of removing foregrounds 
became more essential with the attempts to measure the polarization of the CMB radiation,
especially the enigmatic $B$-modes. 
The latter motivates choice of foregrounds that we deal with in the present paper. 
Synchrotron and dust emissions are the sources of the polarized contamination 
for the CMB polarization studies.   

Diffuse Galactic synchrotron emission is an important polarized foreground source, 
the understanding of which
is essential for CMB studies, especially, in the range of 10-100~GHz.
It is known that the observed spectra of synchrotron emission and synchrotron polarization
(see de Oliveira-Costa \& Tegmark 1999 and references therein)
reveal a range of power laws. 
The polarization of synchrotron emission traces magnetic fields and is perpendicular 
to the plane-of-sky magnetic field direction. 
Since the Galactic synchrotron emissivity is roughly proportional to
the magnetic energy density,
angular spectrum of synchrotron emission reflects statistics of
magnetic field fluctuations in the Galaxy 
(see \S\ref{sect:synchem} for  discussions).

Thermal emission from dust is also an important source of
polarized foreground emission in the range of frequencies larger than 60~GHz. 
The emission gets polarized due to grain alignment (see Lazarian 2007 for a review).   
Therefore, polarization of dust, similar to synchrotron polarization, 
traces magnetic field fluctuations.

What is the cause of the magnetic field fluctuations?
As magnetic field lines are twisted and bend by turbulent motions in the Galaxy 
it is natural to think of the turbulence as the origin of the magnetic field fluctuations.
In fact, several earlier studies addressed the relation between turbulence and
the diffuse synchrotron foreground radiation.
Tegmark et al.~(2000) suggested that the spectra may be relevant to Kolmogorov turbulence.
Chepurnov (1999) and CL02 used different approaches, but both showed that 
the angular spectrum of synchrotron
emission reveals Kolmogorov spectrum ($C_l\propto l^{-11/3}$) for large\footnote{
   In {\it homogeneous} turbulence,
   `large values of multipole' means $l$ larger than $l_{cr}\equiv \pi/\theta$, where
   $\theta$ (in radian) is the angular size of the farthest eddies.
   CL02, for example, discussed that $l_{cr} \sim 30$ for the Galactic halo.
   However, in {\it inhomogeneous} turbulence,
   the practical value for $l_{cr}$ is an order of magnitude larger than $\sim 30$
   (see Chepurnov 1999; CL02). 
} values 
of multipole $l$.
However, they noted that the spectrum can be shallower than the Kolmogorov one for 
smaller values of multipole $l$, due to density stratification
in the halo (Chepurnov 1999) or the Galactic disk geometry (CL02)\footnote{In both approaches larger 
emissivity towards the disk plane is employed.}.

Recent research of compressible magnetized turbulence suggests that the fluctuations may not 
be necessarily Kolmogorov, to start with (see Beresnyak et al. 2005; Kowal \& Lazarian 2007). 
Nevertheless, both observations and numerical studies confirm 
the power-law dependence of turbulence, 
even if the spectrum differs from the Kolmogorov one. 
This is also supported by theory, which states the self-similarity of turbulence. 
It is this self-similarity that gives us hope for a successful removal 
of foregrounds arising from
turbulent interstellar media. 

At the same time, it is known that the spatial spectral index of foregrounds 
measured for different Galactic latitudes may differ. 
If the reason for this is unknown, this may make the weeding out of foregrounds 
using spatial templates unreliable. CL02 identified these changes 
with the variations of the geometry of the emission region. 
Thus, for the same Galactic latitudes one should expect the same slope 
of the spatial spectrum of the fluctuations, which, for instance, 
allows one to extend the power-law spectrum of fluctuations measured 
for low spherical harmonics to higher spherical harmonics. 
Potentially, if the geometry of the emission region is known, this allows us 
to predict the expected changes of the index\footnote{Inverting arguments in CL02 
one can use the changes of the foreground spectra to model the geometry of the emitting volume.}. 
In this paper we provide more support for the conjecture in CL02.  
 
In this paper, we first present general properties and structure functions
of a publicly available synchrotron foreground
emission map.
Then we investigate what kinds of Galactic halo structures can produce
the observed structure function (and therefore angular spectrum), which will be useful
for the study of polarized synchrotron foreground.
We also present the properties
of a publicly available model dust emission map.
Dust emission is one of the most important
sources of polarized foreground radiation.
Therefore, measurement of angular power spectrum of such foreground is
of great interest.
Thus, we provide estimation of
angular spectrum of polarized 
emission by foreground dust.
This result is of great importance in view of recent interest to
the CMB polarization. In \S 2 we explain a way of spatial removal of
foreground emission and provide the summary of the expected scaling
of foreground fluctuations arising from Galactic turbulence.
In \S3, we present statistical analysis of the Haslam map, which
   is dominated by diffuse Galactic synchrotron emission.
In \S4, we investigate polarized emission from thermal dust.
In \S5, we discuss how to utilize our knowledge
 to remove Galactic foregrounds. We provide the discussion of our
results in \S6 and the summary in \S7.
In Appendix, we review a simple model of the angular spectrum of 
synchrotron emission arising from MHD turbulence.
In Appendix, we also present calculations of high-order structure functions of
 the synchrotron and the dust maps and we compare the results with
 those of turbulence.

\section{Motivation: a new technique of foreground removal}

\subsection{Spatial Removal of Foregrounds} \label{sect_2new}

Let us illustrate a possible procedure of the removal of Galactic 
foregrounds from the CMB signal.
The cosmic microwave signals consists of the CMB 
signal $I^{CMB}$ and foregrounds $I^F$. 
When we correlate the microwave signal at points ``1'' and ``2'', we get
\begin{equation}  \langle(I^{CMB}_1+I^F_1)(I^{CMB}_2+I^F_2)\rangle 
    = \langle I^{CMB}_1I^{CMB}_2\rangle+\langle I^F_1 I^F_2\rangle,
\end{equation}
where we assume $I^{CMB}$ and $I^F$ are uncorrelated.
Therefore, the measured angular spectrum $C^{measured}_l$ is 
just the sum of $C^{CMB}_l$ and $C^F_l$ and we have 
\begin{equation}  C^{CMB}_l=C^{measured}_l-C^F_l.
\label{filter}
\end{equation}
That is, if we know the angular spectrum of foregrounds $C^F_l$, 
we can obtain the CMB angular spectrum $C^{CMB}_l$.

How can one obtain $C^F_l$? The well tested way of doing this is to use 
multi-frequency measurements of the CMB + foreground emission and separate 
the two components using the frequency templates of foregrounds. 
This approach requires many measurements at different frequencies. 
In addition, for some foregrounds the frequency templates may be difficult to obtain. 
The so-called "spinning dust" foreground introduced in Draine \& Lazarian (1998ab) 
presents an example of such a difficult-to-remove foreground. 
We also mention that the measurements of foregrounds at different 
frequencies may have different spatial resolutions and the use of 
the maps with different resolutions may also present a problem.

In this paper we address a somewhat different problem, 
which in its extreme\footnote{Less extreme cases would involve the use 
of the known spatial properties of $C^F_l$ to increase the accuracy 
of the removal of foregrounds within traditional techniques. 
We do not discuss these more sophisticated procedures in this paper.} 
can be formulated in the following way. Imagine that we separated the 
foreground and the CMB signals at low resolution $l_{low}$ using the 
traditional multifrequency approach. Is it possible to use this 
information to remove the foreground contribution from $C^{measured}_l$ for $l>l_{low}$? 
For instance, the measurements of the Wilkinson Microwave Anisotropy Probe (WMAP) 
provide a high accuracy measure of $C^F_l$ 
over a limited range of scales. If we know $C^F_l$ as a function of $l_{low}$ 
at scales smaller than those measured by the WMAP, then one can extrapolate $C^F_l$ to $l>l_{low}$. 
These values of $C^F_l$ can be used to filter the microwave 
measurements from balloon-born experiments using the procedure given by Eq.~(\ref{filter}). 
Note that the balloon-born experiments usually have higher spatial resolutions, 
but not enough frequency coverage to remove foregrounds using the frequency templates. 

The key question is to what extend we can predict $C^F_l$ over a range of scales 
which is different from the range of scales at which  $C^F_l$ was measured. 
The answer to this question is trivial if $C^F_l$ is a simple power law. 
While the actual spectra of foregrounds are more complex, in this 
paper we provide both theoretical arguments and the analysis of the 
foreground data which support the notion that $C^F_l$ can be successfully 
extended beyond the range of $l$ which is measured.  

We should stress that the filtering above is different from the accepted techniques 
of foreground removal using frequency-dependent templates. 
The outcome of the latter procedures are maps of the foreground radiation and the CMB radiation. 
The filtering described above is of statistical nature. 
The result of it is $C^{CMB}_l$ rather than emission maps\footnote{It is easy to see that 
the phase information of foreground emission is lost in the process of such a filtering. 
However this information is not necessary for the $C^{CMB}_l$ recovery.}.

In view of above, it is important to determine to what extend the spatial properties 
of $C^F_l$ are predictable, in particular, explore to what extend the power-law 
approximation is applicable. This paper provides a study of spatial statistical 
properties of fluctuations of synchrotron and dust emission and relates those to 
the properties of the underlying turbulence. It also provides an example of the 
filtration procedure that we advocate.  

Another important issue is to determine the reasons for the change of the power 
law behavior with latitude. This is what we study below for the synchrotron and 
dust foreground emissions. The ultimate goal of this research is to obtain models 
of Galactic foregrounds which provide a good fit for the foreground spatial spectrum 
at arbitrary scales and arbitrary latitudes. This paper is a step to constructing such a model. 

\subsection{Power-law behavior of interstellar turbulence}
For the spatial removal procedure to work, we should understand
the spatial spectra of foregrounds.
Since the spatial fluctuations of foregrounds inherit 
the spectra of the underlying interstellar turbulence, we summarize
the spectral behavior of interstellar turbulence.

It is generally accepted that the interstellar medium (ISM) is magnetized and turbulent 
(see reviews by Elmegreen \& Scalo 2004 and McKee \& Ostriker 2007). 
The so-called "Big Power Law in the Sky" corresponding to the Kolmogorov-type turbulence was  
reported in Armstrong, Rickett, \& Spangler (1995). 
Recently this law based on the measurements of the radio scintillations 
arising from electron density inhomogeneities has been extended to larger scales 
using the WHAM H$_\alpha$ fluctuations (Chepurnov \& Lazarian 2010).

Spectra of magnetic turbulence has been studied using Faraday rotation measurements 
(see Haverkorn et al. 2008) and starlight polarization (Hildebrand et al. 2009). 
The interpretation of the results are more challenging in those cases.  

Molecular and atomic spectral lines present a very promising way of studying turbulence. 
The interstellar lines are known to be Doppler-broadened due to turbulent motions. 
Obtaining spectra from Doppler-broadened lines is not trivial, however. 
A lot of research in this direction based on the use of the so-called "velocity centroids", 
which were the main tool to study velocity fluctuations, has been shown to produce erroneous 
results for supersonic turbulence (Lazarian \& Esquivel 2003; Esquivel \& Lazarian 2005; Esquivel et al. 2007). 
At the same time new techniques based on the theoretical description of the 
Position-Position-Velocity (PPV) data cubes, namely, the Velocity Channel Analysis (VCA) 
and the Velocity Coordinate Spectrum (VCS) have been developed 
(Lazarian \& Pogosyan 2000, 2004, 2006, 2008), 
tested (see Stanimirovic \& Lazarian 2001; Lazarian, Pogosyan, \& Esquivel 2002; Esquivel et al. 2003; 
Chepurnov \& Lazarian 2009; Padoan et al. 2006, 2009) 
and applied to the observational data to obtain the characteristics of the velocity  turbulence 
(see Lazarian 2009 for a review). 

VCA and the VCS techniques reveal that the velocities for interstellar turbulence 
may be somewhat steeper than the Kolmogorov one , 
while the density spectra of the fluctuations may be substantially 
more shallow than the spectrum of Kolmogorov fluctuations 
(see Padoan et al. 2006, 2009; Chepurnov et al. 2010). 
This agrees well with the numerical studies of the magnetized supersonic turbulence 
(Beresnyak, Lazarian \& Cho 2005; Kowal, Lazarian \& Beresnyak 2007; Kowal \& Lazarian 2010). 
For the subsonic turbulence the spectrum of magnetized media gets 
the values close to the Kolmogorov index 
(see Goldreich \& Sridhar 1995; Lazarian \& Vishniac 1999; Cho \& Vishniac 2000; 
Muller \& Biskamp 2000; Maron \& Goldreich 2001;  Lithwick \& Goldreich 2001; 
Cho, Lazarian \& Vishniac 2002; Cho \& Lazarian 2002b, 2003; 
Boldyrev 2006; Beresnyak \& Lazarian 2006, 2009).    
In view of that we believe that the Kolmogorov spectrum can be used as 
a proxy of the underlying spectra of velocity and magnetic field, 
while a more cautious approach should be demonstrated to density 
in highly compressible environments, e.g. molecular clouds. 

Below we shall show that the spectra of turbulence derived from the analysis 
of foreground is consistent with both the results of dedicated observations 
of turbulence as well as the theoretical expectation for the MHD turbulence.

\subsection{Data Sets}
We use the 408MHz Haslam all-sky map (Haslam et al.~1982)
and a model 94GHz dust emission map 
that are available on the NASA's LAMBDA website \footnote{
          http://lambda.gsfc.nasa.gov/}.
Both maps were reprocessed for HEALPix (G\'{o}rski et al. 2005)
with nside=512 (7$^\prime$ resolution).

The original Haslam data were produced by merging several different data-sets.
``The original data were 
processed in both the Fourier and spatial domains to mitigate baseline 
striping and strong point sources'' (see the website for details). 
The angular resolution of the original Haslam map is $\sim 1^{\circ}$.
Galactic diffuse synchrotron emission is the dominant source of
emission at 408MHz.

The 94 GHz dust emission map 
is based on the work of Schlegel, Finkbeiner, \& Davis (1998) and
Finkbeiner, Davis, \& Schlegel (1999).
Schlegel et al. (1998) combined 100$\mu m$ maps of
IRAS (Infrared Astronomy Satellite) and 
DIRBE (Diffuse Infrared Background
Experiment on board the COBE satellite) and removed the
zodiacal foreground and point sources to
construct a full-sky map.
Finkbeiner et al. (1999) extrapolated the 100$\mu m$ emission
map and 100/240$\mu m$ flux ratio maps to sub-millimeter and
microwave wavelengths.
The 94 GHz dust map we used is identical to the two-component model 8 of 
Finkbeiner et al. (1999).
The angular resolution of the 100-micron map is $\sim 6^{\prime}$ and
that of the temperature correction derived from the 100/240-micron ratio map
is $\sim 1^{\circ}$ (Finkbeiner et al. 1999),
which corresponds to $l\sim 180^\circ/\theta^\circ \sim 180$.



\section{Spatial statistics of Diffuse Galactic Synchrotron Emission}
In this section, we analyze the Haslam 408MHz all-sky map, which is
dominated by Galactic diffuse synchrotron emission.
Our main goal is to explain the observed synchrotron angular spectrum
using simple turbulence models.
The result in this section will be useful for sophisticated modeling of
polarized synchrotron emission.

\subsection{General properties of diffuse Galactic synchrotron emission}


In this section, we study synchrotron emission from the Galactic halo 
(i.e. $b \gtrsim 30^{\circ}$) and the Galactic disk (i.e. $|b|\le 2^{\circ}$)
separately. 
Our main goal is to see if statistics of synchrotron emission from the halo
is consistent with turbulence models.
When it comes to synchrotron emission from the Galactic disk,
it is not easy to separate diffuse emission and emission from discrete sources.
Therefore, we do not try to study turbulence in the Galactic disk.
Instead, we will try to estimate which kind of emission is dominant
in the Galactic disk.

There exist several models for the diffuse Galactic radio emission.
Beuermann, Kanbach, \& Berkhuijen (1985) showed that a two-component model,
a thin disk embedded in a thick disk, can explain observed synchrotron 
latitude profile.
They claimed that the equivalent width of the thick disk is 
about several kiloparsecs
and thin disk has approximately the same equivalent width as the gas disk.
They assumed that, in the direction perpendicular to the Galactic plane,
the emissivity $\epsilon$ of each component follows 
\be
   \epsilon(z) = \epsilon(0) sech^b(z/z_0),
\ee
where $z$ is the distance from the Galactic plane 
and $\epsilon(0)$, $b$, and $z_0$ are constants.
The half-equivalent width of the disk, which
is proportional to $z_0$, at the location of the Sun is $\sim 2$kpc.

Recently, several Galactic synchrotron emission models have been proposed
in an effort to separate Galactic components from the WMAP polarization data
(see, for example, 
                   Page et al.~2007; 
                   Sun et al.~2008; Miville-Deschenes et al.~2008; 
                   Waelkens et al.~2008).
All the models mentioned above assume the 
existence of a thick disk component with
a scale height equal to $1~kpc$.
Sun et al.~(2008) considered an additional local spherical component
motivated by the local excess of the synchrotron emission that might be related to
the ``local bubble'' (see, for example, Fuchs et al. 2008).

The detailed modeling of the Galactic synchrotron emission is beyond the 
scope of our paper. 
We will simply assume that there is a thick component with a scale height of
$\sim 1$kpc. We will also assume that 
there could be an additional local spherical component.
Then, in the following subsections,
we will consider the relation between spectrum of 3-dimensional turbulence 
and the observed angular spectrum of synchrotron emission.



\begin{figure*}[h!t]
\includegraphics[width=0.45\textwidth]{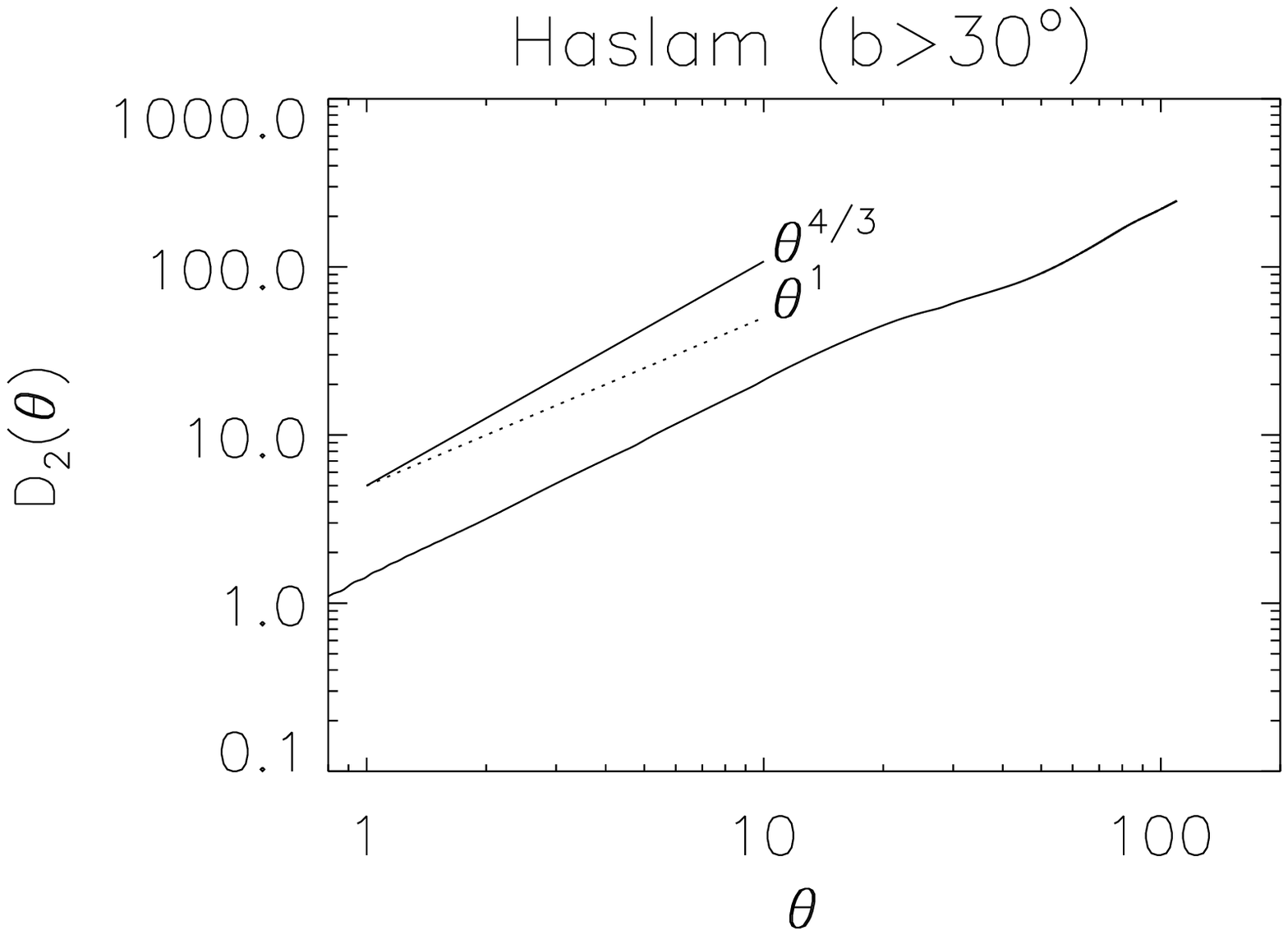}   
\includegraphics[width=0.45\textwidth]{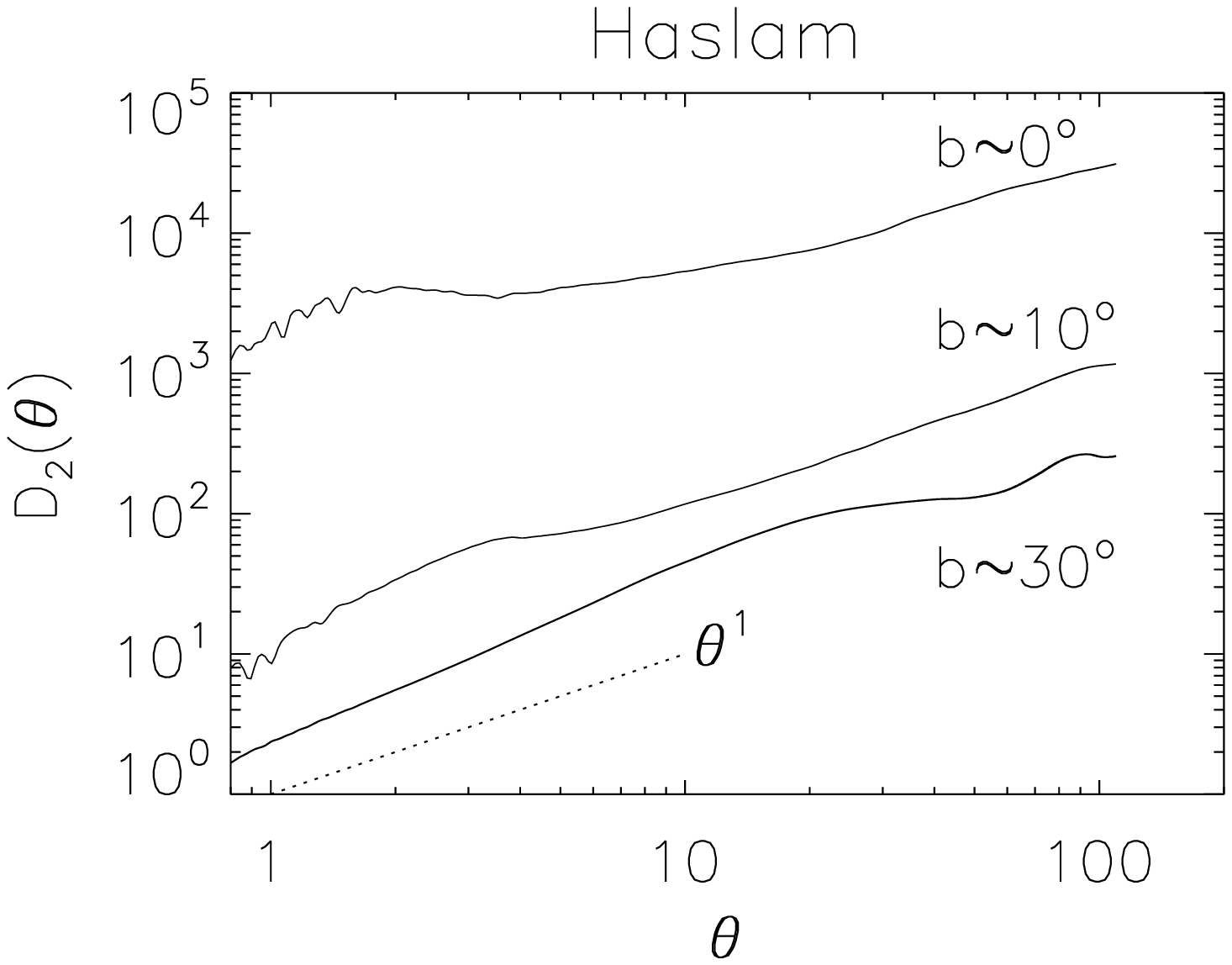}   
\caption{ 
   Haslam 408MHz map.
 {\it Left}: Second-order angular 
   structure function (for $b>30^{\circ}$) shows a slope of $\sim 1.2$, which
   is shallower than that of Kolmogorov turbulence (slope $= 5/3$).
  {\it Right}: Structure function as a function of Galactic latitude.
    {}From bottom to top, the second-order angular 
     structure functions are obtained
     for thin stripes ($|\Delta b| \le 2^\circ$) along 
     Galactic latitudes of $30^{\circ}$,
      $10^{\circ}$, and $0^{\circ}$.
}
\label{fig:sfetc-has}
\end{figure*}

\subsection{Structure function of the 408MHz Haslam map}
In Appendix,  
we discussed the relation between
the 3D spatial MHD turbulence spectrum and the observed 2D angular spectrum
of synchrotron emission (see Eq.~[\ref{eq_1_11_3}] for a quick summary).
But, in Appendix we assumed that the emission arises from a spherical
region filled with homogeneous turbulence.
In this section, we will show that
the modulation of synchrotron intensity of the emitting volume can also affect
the observed angular spectrum of synchrotron emission.


Earlier studies showed that the angular spectrum 408-MHz Haslam map
has a slope close to $-3$: $C_l \propto l^{-3}$
(Tegmark \& Efstathiou 1996; Bouchet, Gispert, \& Puget 1996).
Recently La Porta et al. (2008) performed a 
comprehensive angular power spectrum
analysis of all-sky total intensity maps at 408MHz and 1420MHz. 
They found that the slope is close to -3 for high Galactic latitude regions.
Other results also show slopes close to -3. For example, 
using Rhodes/HartRAO data at 2326 MHz (Jonas, Baart, \&
Nicolson 1998), Giardino et al. (2001b) obtained a slope $\sim 2.92$ 
for high Galactic latitude regions with
$|b| > 20^\circ$. Giardino et al. (2001a) obtained a slope $\sim 3.15$ for high
Galactic latitude regions with $|b| > 20^\circ$ from the Reich \& Reich
(1986) survey at 1420 MHz. Bouchet \& Gispert (1999) 
also obtained a slope $\sim l^{-3}$
spectrum from the 1420 MHz map.

%
%

In general, synchrotron emission from the Galactic disk makes
it difficult to measure the angular spectrum of synchrotron emission from
the Galactic halo.
In order to avoid the contamination by the Galactic disk, one may mask out the
low Galactic latitude regions.
This can be done, for example, by setting all synchrotron intensity to zero
for regions with $|b| < b_{cut}$.
However, the angular spectrum obtained with the Galactic mask 
exhibits spurious oscillations. 
Moreover, the spectrum obtained with a mask may not be the true one
because it is contaminated by the mask.
In principle, one may correct such oscillations and estimate the true spectrum, 
using the convolution theorem:
the Fourier coefficients (or, in this case, spherical harmonic coefficients) of the masked data
are convolution of those of the true data and those of the mask.
However, the practical implementation of the method is not simple.

Another approach is to estimate 2-point correlation function first and 
to extract the angular power spectrum from it (Szapudi et al. 2001; see also Eq.~[\ref{eq:k2cl}]).
This method is free of artifacts caused by the mask.
%
%
But the angular spectrum $C_l$ obtained in this way is, in general, noisy and
it requires a lot of calculations to accurately measure the slope of the spectrum.

We are interested in the slope of the angular spectrum
on small angular scales, which is the same as
that of the underlying 3D spatial turbulence spectrum (see Appendix).
Therefore, in this paper, we use yet another approach.
We first calculate
the second-order angular structure function:
\be
   D_2 (\theta) 
   = < | I({\bf e}_1) -I({\bf e}_2) |^2 >, 
\ee
where $I({\bf e})$ is the intensity of synchrotron emission,
${\bf e}_1$ and ${\bf e}_2$ are unit vectors along the lines of
sight, $\theta$ is the angle between ${\bf e}_1$ and ${\bf e}_2$, and
the angle brackets denote average taken over the observed region.
Then, we extract the slope of the angular spectrum using the 
relation
\be
    D_2(\theta) \propto \theta^{m-2} \Leftrightarrow 
   C_l \propto l^{-m}   \Leftrightarrow E_{3D}\propto k^{-m}
\ee
for small angular scales (see \S\ref{sect:smangle}).

We note that excessive care is required in the presence of white noise.
 In the presence of white noise, second-order structure function will be
 $D_2 (\theta) 
   = < | I({\bf e}_1) +\delta_1 -I({\bf e}_2)-\delta_2 |^2 >
  = < | I({\bf e}_1) -I({\bf e}_2) |^2 > + < |\delta_1-\delta_2|^2 >,
  $
where $\delta_1$ and $\delta_2$ represent noise (Chepurnov, private communication).
If the second term on the right ($< |\delta_1-\delta_2|^2 >$) is
sufficiently smaller than the first term on the right ($< | I({\bf e}_1) -I({\bf e}_2) |^2 >$),
we can ignore the noise.
Otherwise,
the noise can interfere accurate measurement of the slope.
In the Haslam map, the level of the white noise seems to be negligible.
The reason is as follows.
Our measurements show that $D_2(0.015^\circ) \sim 0.05$ in the Haslam map.
This means that the $< |\delta_1-\delta_2|^2 >$ term is no larger than
0.05, which is sufficiently smaller
than values $D_2(\theta)$ shown in Fig.~\ref{fig:sfetc-has}.


In the left panel of Fig.~\ref{fig:sfetc-has} we show the second-order
structure function for the Galactic halo (i.e. $|b| > 30^{\circ}$).
The slope of the second-order structure function lies between
those of two straight lines.
The steeper line has a slope of 4/3 and the other one has a slope of 1.
The actual measured slope is $\sim 1.2$.
This result implies that the 3D turbulence spectrum is $E_{3D}(k)\propto k^{-3.2}$,
which is shallower than the Kolmogorov spectrum $E_{3D}(k)\propto k^{-11/3}$.

Now a question arises: why is the slope shallower than that of Kolmogorov?
Chepurnov (1999) provided a discussion 
of the effects of density stratification on the slope.
He used a Gaussian disk model and semi-analytically showed that
the slope of the angular spectrum can be shallower than
that of Kolmogorov. In the next subsection, we present our understanding of the effect. 

\begin{figure}[h!t]
\includegraphics[width=0.40\textwidth]{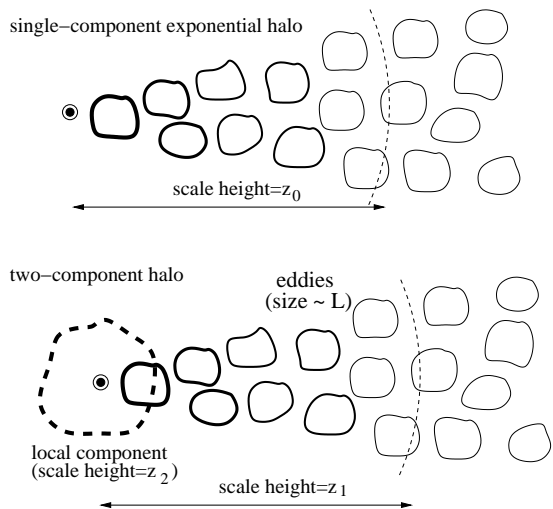}  
\caption{Halo models with stratification.
  {\it Upper plot}:
  Exponentially stratified halo.
  We take $z_0 (=r_0) =1kpc$ and $L$ (=eddy size) $=100pc$.
{\it Lower plot}: Two-component halo.
  We take  $z_1 (=r_1) =1kpc$, $z_2 (=r_2) =100pc$,  
  and $L$ (=eddy size) $=100pc$.
}
\label{fig:2models}
\end{figure}

\subsection{Model calculations} 
\label{sect:3models}

 Below we take into account that the emission does depend on
the distance from the Galactic plane.
Synchrotron emission models (see previous subsection) assume either exponential 
($e^{-z/z_0}$) or square of hyperbolic secant ($\mbox{sech}^b[z/z_0]$) law
for synchrotron emissivity, where $z$ is the distance from the 
Galactic plane and $b$ is a constant.
For simplicity, we assume the observer is at the center of
a spherical halo. That is, the geometry is not plane-parallel, but spherical.
In what follows, we use $r$, instead of $z$, to denote the 
distance to a point.

To illustrate the effects of this inhomogeneity, we test 3 models:
\begin{enumerate}
\item {\bf Homogeneous halo}: Turbulence in halo, thus emissivity, is homogeneous. 
      Turbulence has a sharp
      boundary at $d_{max}=1kpc$. The outer scale of turbulence is 100pc.
      Basically, this model is the same as the one we considered
      in \S\ref{sect:spsfco}.   
\item {\bf Exponentially stratified halo}: Emissivity shows an exponential decrease, 
      $\epsilon(r) \propto e^{-r/r_0}$.
      We assume $r_0 = 1kpc$ and the halo truncates at $r=8kpc$.
      The outer scale of turbulence is 100pc. See Fig.~\ref{fig:2models}.
\item {\bf Two-component halo}: Emissivity decrease as 
      $\epsilon(r) \propto \epsilon_1 e^{-r/r_1}+\epsilon_2 e^{-r/r_2}$, where
      $\epsilon_2=10\epsilon_1$, $r_1 = 1kpc$, and $r_2=100pc$.
      The halo truncates at $r=8kpc$.
      The outer scale of turbulence is 100pc.
      The second component mimics local enhancement of synchrotron emissivity. See Fig.~\ref{fig:2models}.
\end{enumerate}

\begin{figure*}[h!t]
\includegraphics[width=0.32\textwidth]{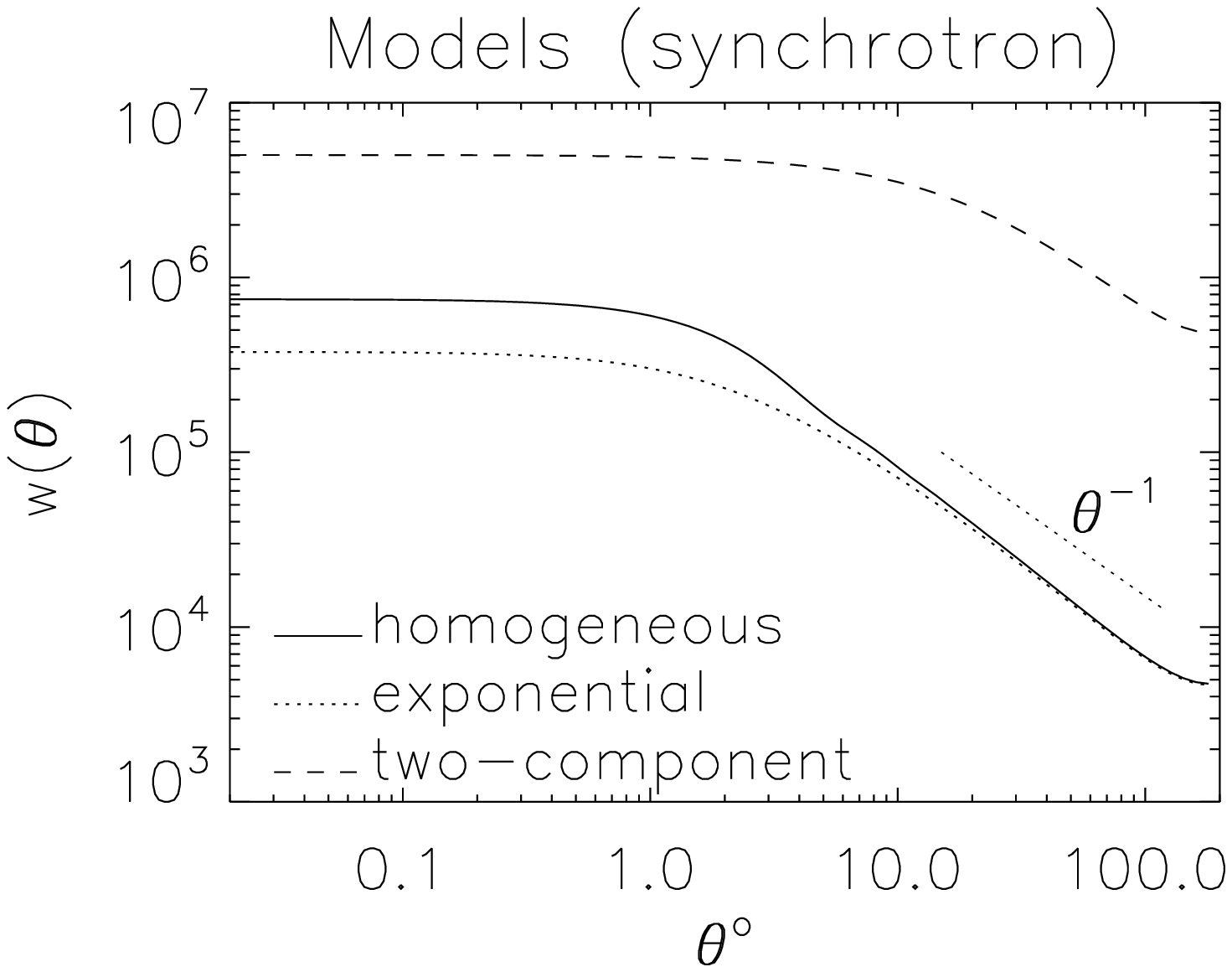}  
\includegraphics[width=0.32\textwidth]{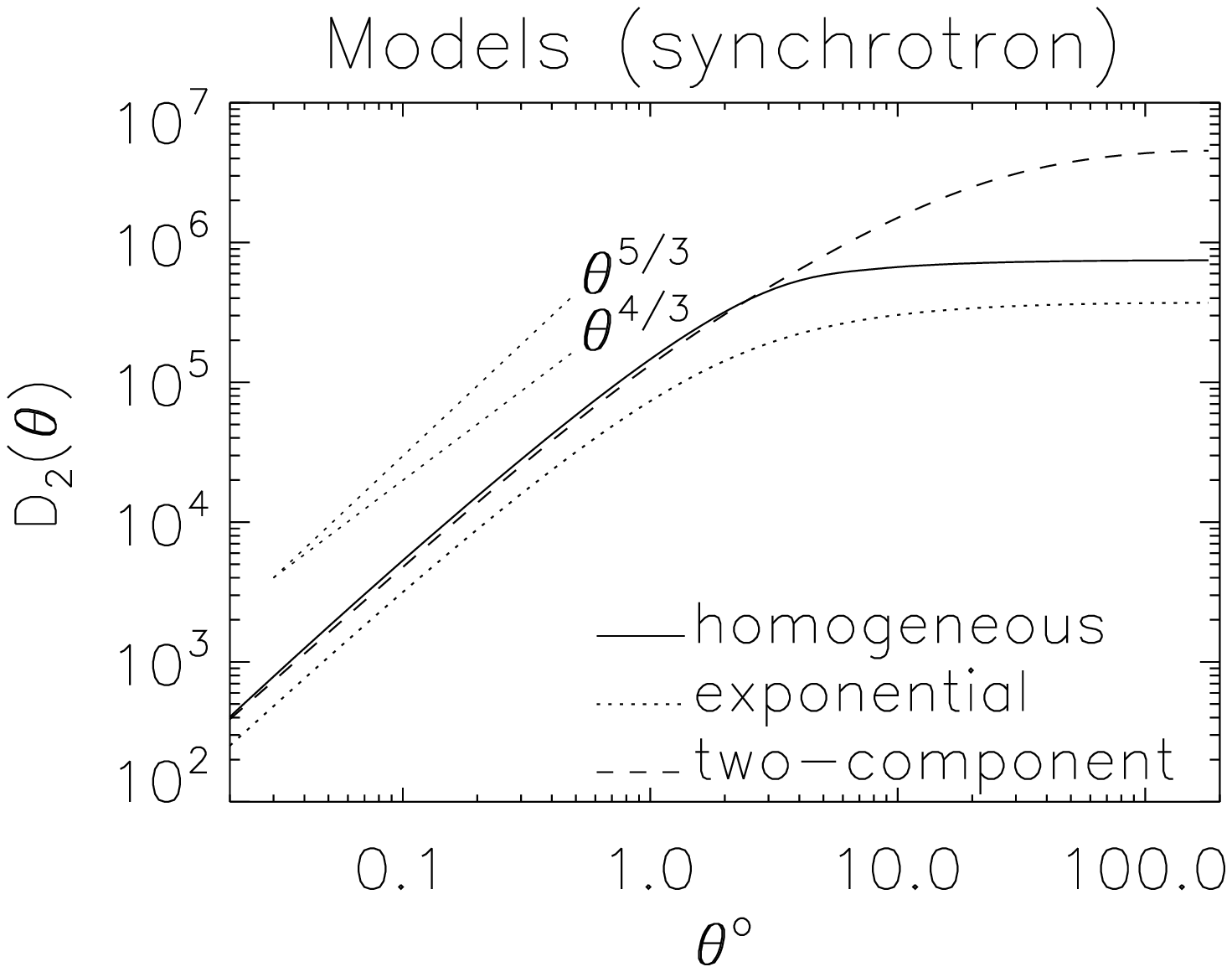}   
\includegraphics[width=0.32\textwidth]{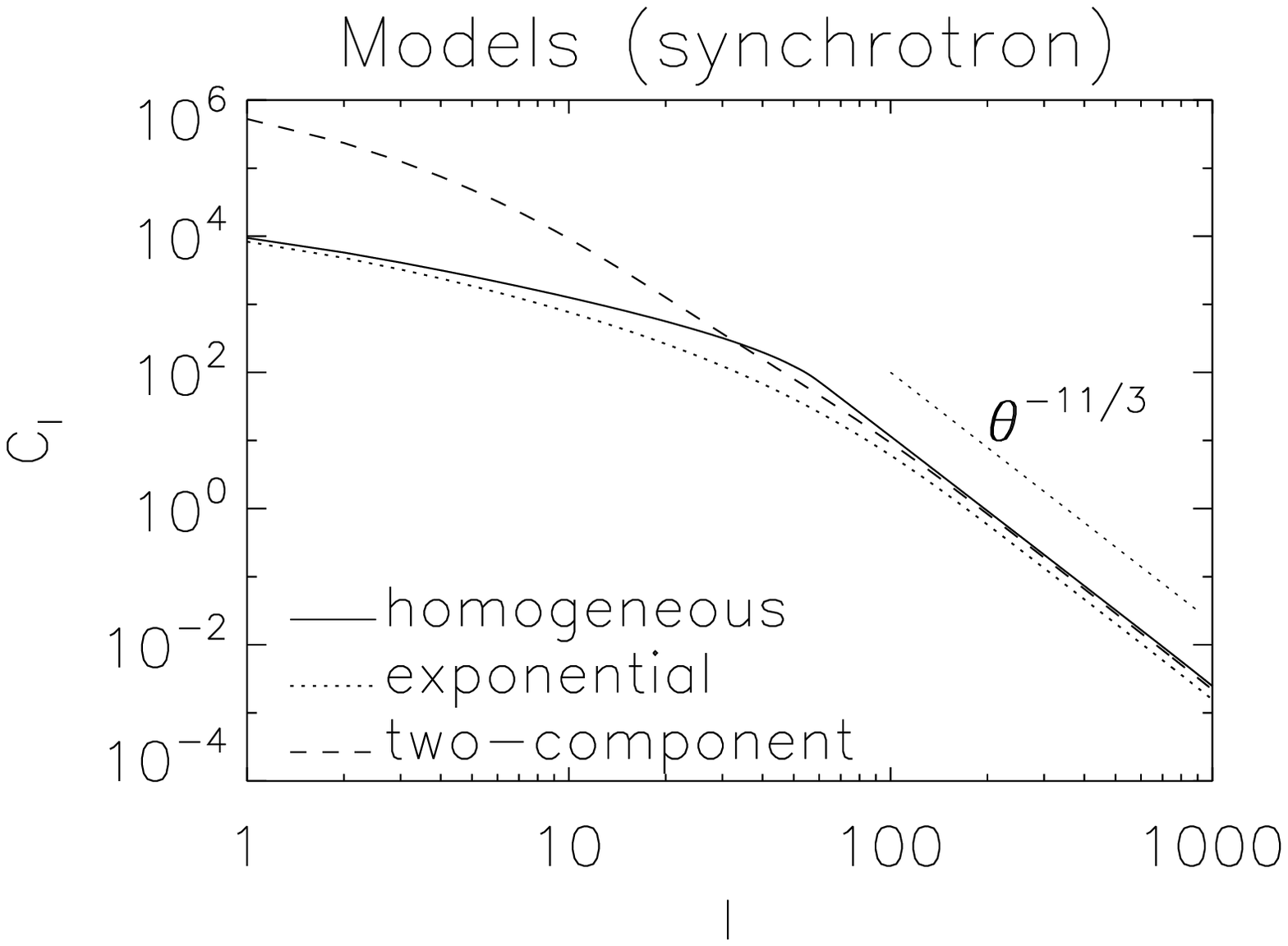}  
\caption{ 
   Model calculations. 
   Three toy models for emissivity profiles in the Galactic halo are considered:
   homogeneous (solid line) halo, exponentially stratified halo (dotted line),
   and two-component exponential halo (dashed line).
      {\it Left}: Angular correlation functions.
                     When angular separation is large, angular correlation
                     functions follow the universal relation:
                         $(\pi -\theta)/\sin\theta \sim 1/\theta$.
      {\it Middle}: Second-order structure functions.
                    When angular separation is small,
                    the slope of the homogeneous turbulence (solid line) 
                    is compatible with $5/3$.
                    But those of 
                    stratified halo models are shallower.
                    The two-component model (dashed line) shows a slope
                    compatible with $-3$, if we measure average slope
                    between $\theta \sim 0.5^\circ$ and $\sim 10^\circ$.
      {\it Right}: Angular spectra.
                   All spectra are compatible with the 3D spatial
                   turbulence spectrum of $l^{-11/3}$ for $l\gtrsim$ a few time 100.
                   The stratified halo models show 
                   shallower slopes, if we naively attempt to fit it with a single power law and measure                        the resulting averaged slope between $l\sim 10$ and $\sim 200$.
                   The homogeneous turbulence model also gives a
                   shallower slope for these values of $l$.
                   But, its spectrum shows a distinct break
                   near $l\sim 50$.
}
\label{fig:model}
\end{figure*}

We numerically calculate the angular correlation function $w(\theta)$ and
the second-order structure $D_2(\theta)$ from
\bea
    w(\theta)=\int dr_1 \int dr_2 ~{\cal K}(|{\bf r}_1-{\bf r}_2|)
    \epsilon(r_1) \epsilon(r_2),   \label{eq_ctheta}  \\
    D_2(\theta) \propto \overline{T} - w(\theta),
\eea
where $|{\bf r}_1-{\bf r}_2|=r_1^2+r_2^2-2r_1r_2 \cos{\theta}$,
$\epsilon(r)$ is the synchrotron emissivity, 
$\overline{T}=\lim_{\theta \rightarrow 0} w(\theta)$,
and 
we use the spatial correlation function ${\cal K}(r)$ obtained from the relation:
\be
 {\cal K}(r) \propto \int_{0}^{\infty} 4\pi k^2 E_{3D}(k) \frac{ \sin kr }{ kr } dk
               \label{c_r_3D}
\ee
where the spatial spectrum of emissivity $E_{1D}$ has the form:
\be
 E_{3D}(k) \propto \left\{ \begin{array}{ll} 
                              \mbox{constant}   & \mbox{if $k\le k_0$} \\
                              (k/k_0)^{-11/3}   & \mbox{if $k\ge k_0$,}
                      \end{array}
              \right. \label{E_3D}
\ee
which is the same as Kolmogorov spectrum for $k\ge k_0$ ($ \sim 1/L$).
The reason we use  a constant spectrum for $k\le k_0$ is explained in Appendix B
(see also Chepurnov 1999).
We obtain the angular spectrum from the relation:
\be
   C_l \propto \int P_l(\cos{\theta}) K(\cos \theta)\ d(\cos\theta), \label{eq:k2cl}
\ee
where $P_l$ is the Legendre polynomial.

In Fig.~\ref{fig:model}, we plot the calculation results.
The angular correlation function $w(\theta)$ does not change much 
when $\theta$ is small, and
follows $\sim (\pi-\theta)/\sin\theta \sim 1/\theta$ when $\theta$ is large.
The critical angle is a few degrees for homogeneous model (thick solid curve)
and single-component exponential model (dotted curve).
As we discussed earlier, the critical angle for homogeneous turbulence is
$\sim (L/d_{max})^{rad}\sim 6^\circ$, where $d_{max}$ ($=1kpc$ in our model) 
is the distance to the farthest eddy.
In Fig.~\ref{fig:model} (left panel) we clearly see that
the slope of $w(\theta)$ changes near $\theta \sim 6^\circ$.
The second-order structure function $D_2(\theta)$ 
also shows a change of slope near the
same critical angle ($\theta\sim 6^\circ$).
In single-component exponential model (dotted curve), 
the value of $d_{max}$ is not important.
Instead, the scale height $z_0$ is a more important quantity, 
which is $1kpc$ in our model.
In left and middle panels of Fig.~\ref{fig:model}, we observe
 that the 
single-component exponential model also show a change of slope 
near $\theta \sim$ a few degrees. 
Therefore, we can interpret that the critical angle
for stratified turbulence is $\sim L/z_0$, instead of $\sim L/d_{max}$

Now, it is time to answer our earlier question of why
the observed slope is shallower than that of Kolmogorov.
Let us take a look at the right panel of Fig.~\ref{fig:model}.
All 3 models show that the slope of $C_l$ is almost Kolmogorov one for
$l\gtrsim 200 > l_{cr}\sim \pi d_{max}/L \sim 30$.
However, if we measure {\it average} slope of $c_l$ between $l=10$ and $200$,
we obtain slopes shallower than Kolmogorov.
The single-component model and the homogeneous model
give similar average slopes around $-3$.
However, the homogeneous model
gives a more abrupt change of slope near $l\sim 30$.
In fact, right panel of Fig.~\ref{fig:model} shows a noticeable break 
 near $l\sim 50$.
The average slope of 
the two-component model gives more or less gradual change of the slope and
the observed slope is very close to $-3$ for a broad range of multipoles $l$.
It is difficult to tell which model is better because the models are
highly simplified.
Nevertheless, the two-component model looks the most promising, which
is not so surprising because two-component model has more degree-of-freedom.

Note that, compared with the two-component model, 
the single-component model shows a more or less sudden change of slope
near $l\sim l_{cr} \sim \pi z_0/L \sim 30$.
Therefore, if the single-component model is correct, the scale height $z_0$ cannot be
much larger than $\sim 10$ times the outer scale of turbulence $L$.
If $z_0$ is much larger than $\sim 10L$, $l_{cr}$ becomes smaller and
we will have almost Kolmogorov slope for $l\gtrsim 10$.
We also note that it is possible that
 3D spatial turbulence spectrum itself can be shallower than
the Kolmogorov one. 
That is, it is possible that 
spectrum of ${\bf B}({\bf r})$, hence that of $B^2({\bf r})$, can be
shallower than the Kolmogorov one.
For example, some recent studies show that strong MHD turbulence can have a $\sim k^{-3.5}$
spectrum, rather than $k^{-11/3}$ (Maron \& Goldreich 2001, Boldyrev 2005; 
Beresnyak \& Lazarian 2006)\footnote{The reason for the spectrum being shallow in simulations is unclear. 
It may also be the result of the limited
dynamical range in the presence of non-locality of MHD turbulence (Beresnyak \& Lazarian 2009).}. 
If this is the case, the observed angular spectrum can be slightly shallower than the Kolmogorov for $l>l_{cr}$. 

 To summarize this subsection,
our simple model calculations imply that $C_l$ will be very close to the underlying spectrum
of magnetic turbulence for large values of $l$ ($l\gtrsim$ a few time 100). 
The corresponding spectral slope is expected to be close to the Kolmogorov one (see \S3).  
For intermediate values of $l$ (e.g.~$10<l<200$), the average slope is shallower than
the Kolmogorov one. Thus our modeling shows the consistency of the observational spectra 
with the expectations. Studies of the observed spectra at higher $l$ may be useful for better testing 
of our predictions.

\subsection{Synchrotron emission from Galactic disk}

In right panel of Fig.~\ref{fig:sfetc-has} we show how the second-order structure
function changes with Galactic latitude.
The lower curve is the second-order angular structure function 
obtained from pixels in the range of $28^{\circ}\le b \le 32^{\circ}$.
The middle and upper curves are the second-order 
angular structure functions
obtained from pixels in the range of $8^{\circ}\le b \le 12^{\circ}$ and
$-2^{\circ}\le b \le 2^{\circ}$, respectively.
The middle and upper curves clearly show break of slopes near
$\theta \sim 3^\circ$ and $\sim 1.5^\circ$, respectively.
When the angular separation is larger than the angle of the break,
the structure function becomes almost flat.
As we move towards the Galactic plane, the sudden changes of slopes happen
at smaller angles.

There are at least two possible causes for the break of slope.
First, a geometric effect can cause it. 
As we discussed in the Appendix, change of slope occurs near
$\theta_c \sim L/d_{max}$.
As we move towards the Galactic plane, the distance to the farthest eddy, $d_{max}$,
will increase. As a result, the critical angle $\theta_c \sim L/d_{max}$ will
decrease.
Therefore, we will have smaller $\theta_c$ towards the Galactic plane.
 This may be what we observe in the right panel of Fig.~\ref{fig:sfetc-has}.
Second, discrete synchrotron sources can cause flattening of the structure
function on angular scales larger than their sizes. 
Although the map we use was reprocessed to remove
strong point sources, there might be unremoved discrete sources.
When filamentary discrete sources dominate synchrotron emission, the 
second-order structure function
will be flat on scales larger than the typical width of the sources.
In reality, both effects may work together.

In view of the variations of the spatial spectral slope of the synchrotron emission 
at low Galactic latitudes, the use of the foreground removal procedure discussed in
\S 2 is more challenging for those latitudes. At the same time, the high-$l$ fluctuations
corresponding high latitudes should be possible to remove reliably with the procedure in \S 2 due 
to the observed regular power-law behavior.    

\subsection{On the polarized synchrotron emission}
Roughly speaking, the shape of
the angular spectrum of polarized synchrotron emission will 
be similar to that of the total intensity at {\it mm} wavelengths.
However, it is expected that at longer wavelengths, Faraday rotation and depolarization effects
 should cause flattening of the angular spectrum, which has been
actually reported
(see de Oliveira-Costa et al. 2003 and references therein).
On the other hand, La Porta et al. (2006) analyzed the new DRAO 1.4GHz 
polarization survey and obtained angular power spectra with
power-law slopes in the range $[-3.0, -2.5]$.
More observations on the polarized synchrotron foreground emission
can be found in Ponthieu et al. (2005),
Giardino et al. (2002), Tucci et al. (2002), Baccigalupi et al. (2001).
In this paper, we do not discuss  the properties of the
polarized synchrotron emission.
Readers may refer to recent models about the polarized synchrotron emission
(Page et al. 2007; Sun et al. 2008; Miville-Deschenes et al. 2008;
 Waelkens et al. 2008).

\section{Polarized Emission from Dust}
Polarized radiation from dust is an important
component of Galactic foreground that strongly interferes with the
intended CMB polarization measurements (see Lazarian \& Prunet 2001).
Therefore, the angular spectrum of the polarized radiation from the
foreground dust is of great interest.
One of the possible ways to estimate the polarized dust radiation at the microwave
range is to measure star-light polarization and use
the standard formulae (see, for example, Hildebrand et al. 1999) relating polarization at different wavelengths. This approach involves a number of assumptions the accuracy of which 
we analyze below. In this section, we describe how we can obtain a map of 
polarized dust emission using starlight polarization and we discuss
the angular spectrum of the polarized foreground emission from
thermal dust at the high Galactic latitude (say, $|b| \gtrsim 20^\circ$).

\subsection{Properties of the 94GHz Dust Emission Map} \label{sect:dust_em}



Let us begin with a model dust emission map created by
Finkbeiner et al. (1999), which is 
available at the NASA LAMBDA website (http://lambda.gsfc.nasa.gov/).
As we will explain later in this section, we can derive a polarized
intensity map from this kind of total intensity map.

\begin{figure*}[h!t]
\includegraphics[width=0.45\textwidth]{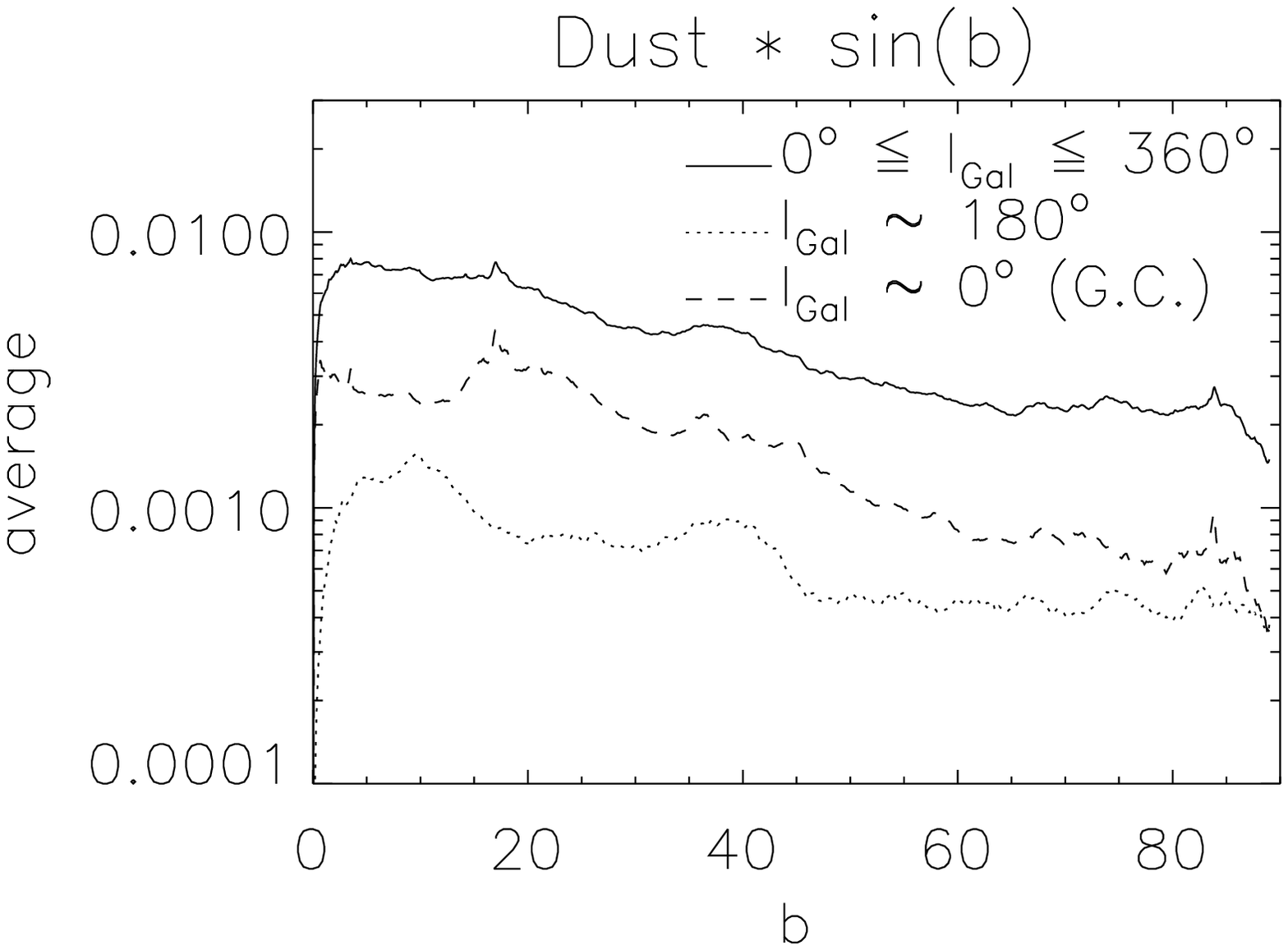}    
\includegraphics[width=0.45\textwidth]{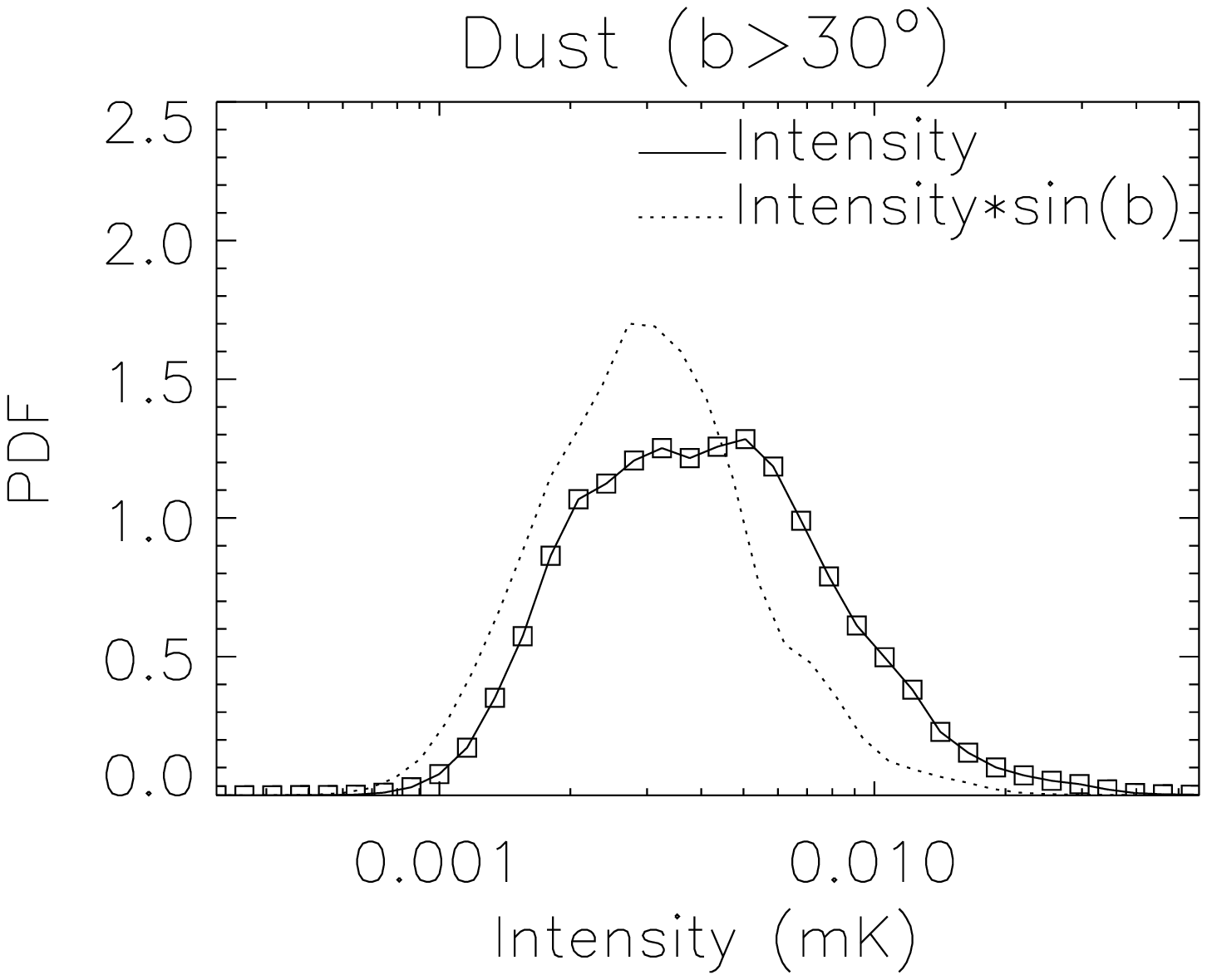}  
\caption{ 
   Average dust emission intensity times $\sin b$. 
   The 94GHz dust emission map has been used.
  {\it Left}: Latitude profile of averaged dust emission 
     intensity time $\sin b$. 
     Average value is roughly constant in high latitude region.
     The solid line represents the average taken over $360^{\circ}$. 
   The dashed line and the dotted line depict average taken near
   the Galactic center ($-45^{\circ}\le l_{Gal} \le 45^{\circ}$) 
   and the Galactic anti-center ($135^{\circ}\le l_{Gal} \le 225^{\circ}$),
   respectively.
 {\it Right}: 
    PDF for $|b|>30^\circ$.  PDF of dust intensity times $\sin b$ 
   (dotted curve) shows a rough symmetry, but
    that of dust intensity (solid curve) does not.
}
\label{fig:dust}
\end{figure*}

We note that the difference between the dust emission and synchrotron emission (discussed in the previous section) is expected. The origin of synchrotron emission is related to cosmic ray electrons which are distributed within an extended magnetic halo (Ginzburg \& Ptuskin 1976). At the same time, dust is expected to be localized mostly within the Galactic plane.
In Fig.~\ref{fig:dust}, we present statistical properties of the map.
The map shows rough constancy of emission
for high Galactic latitude region when
multiplied by $\sin b$ (left panel of Fig.~\ref{fig:dust}).
The $\sin b$ factor also appears in the PDF: the $\sin b$ factor makes the PDF more
symmetric (right panel of Fig.~\ref{fig:dust}).
Therefore it is natural to conclude that a disk component dominates the dust map.

As in the Haslam map, we do not try to obtain angular spectrum of dust emission map
directly.
Instead, we use the second-order structure function in order to reveal the
angular spectrum on small angular scales.
Indeed, the second-order structure function 
of the dust map (see \S\ref{sect:h_o}) shows
a slope of $\sim 0.6$, which corresponds to angular spectrum of $\sim l^{-2.6}$.

The slope of the angular power spectrum of the model dust emission map
is very similar to that of the original FIR data.
Schlegel et al.~(1998) found a slope of $-2.5$ for the original FIR
data. On the other hand, other researchers found
slopes close to $-3$ from other observations 
(see Tegmark et al.~2000 and references therein;
see also Masi et al.~2001).

Dust density fluctuations mostly arise from cold dense phases of interstellar medium 
(see Draine \& Lazarian 1998 for a list of idealized ISM phases). 
There the turbulence is known to be supersonic, which is vividly revealed, for instance, 
by Doppler broadening of observed molecular lines from Giant Molecular Clouds (GMCs) 
(see McKee \& Ostriker 2007).  
The shallow spectrum of density fluctuations is consistent with the numerical simulations of 
supersonic MHD turbulence in Beresnyak et al. (2005). The shallow spectrum was later also 
reported in supersonic hydro turbulence in Kim \& Ryu (2005), which indicates that the effect is not 
radically changed by the magnetic field. The latter makes the conclusion about the shallow spectrum 
independent on the degree of the interstellar medium magnetization and sub-Alfvenic versus 
super-Alfvenic character of turbulence there. Thus we claims the observed spectra should be 
associated with the shallow spectra of underlying density fluctuations in the denser part 
of the interstellar medium. We predict that the shallow power law arising from density 
fluctuations extends from the scales of the turbulence energy injection to the dissipation scales. 
As a result, power-law extension of the observed data and the corresponding filtering 
using Eq.~(\ref{filter}) is possible.   

\begin{figure*}[h!t]
\includegraphics[width=0.45\textwidth]{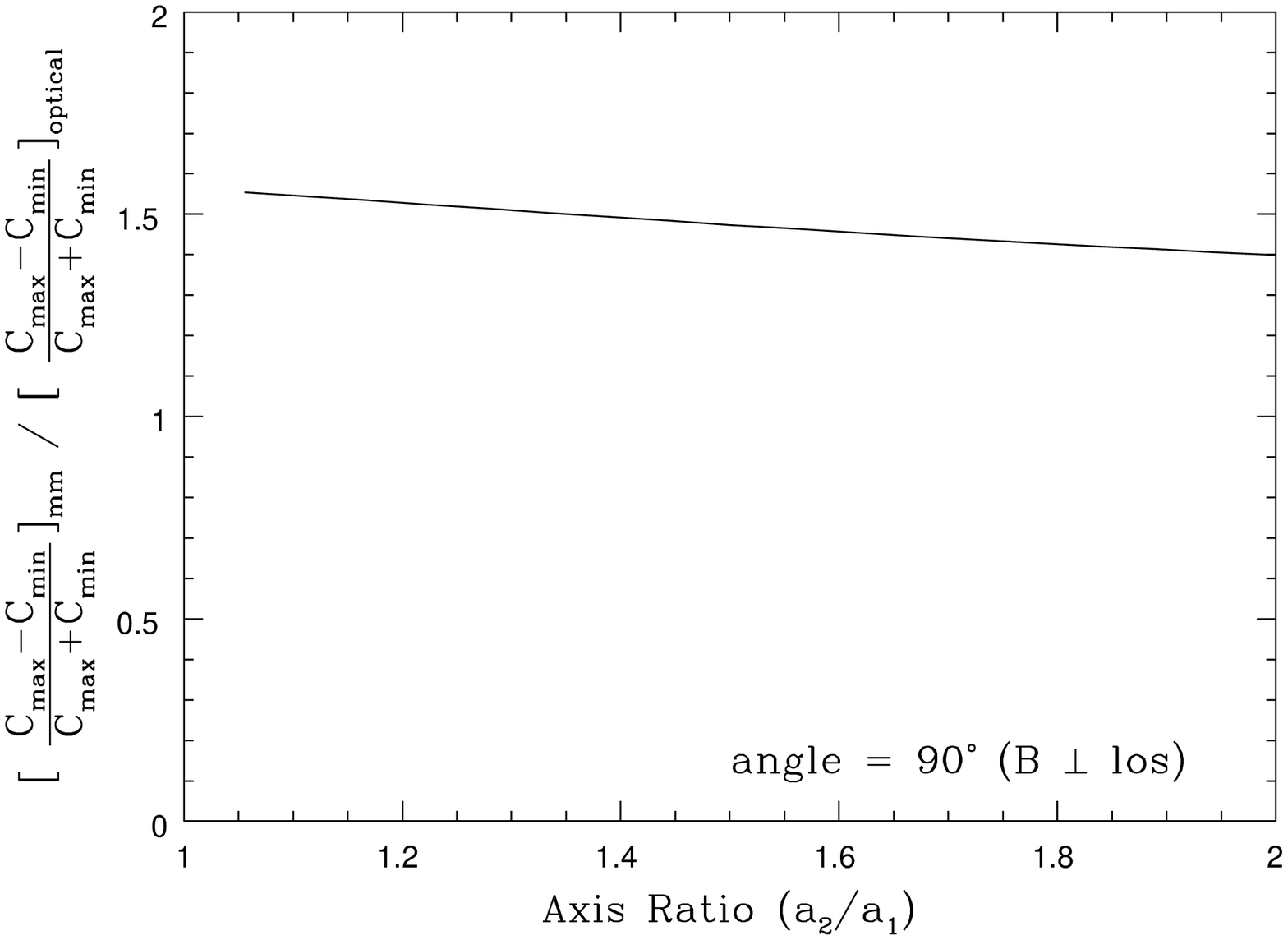}  
\includegraphics[width=0.45\textwidth]{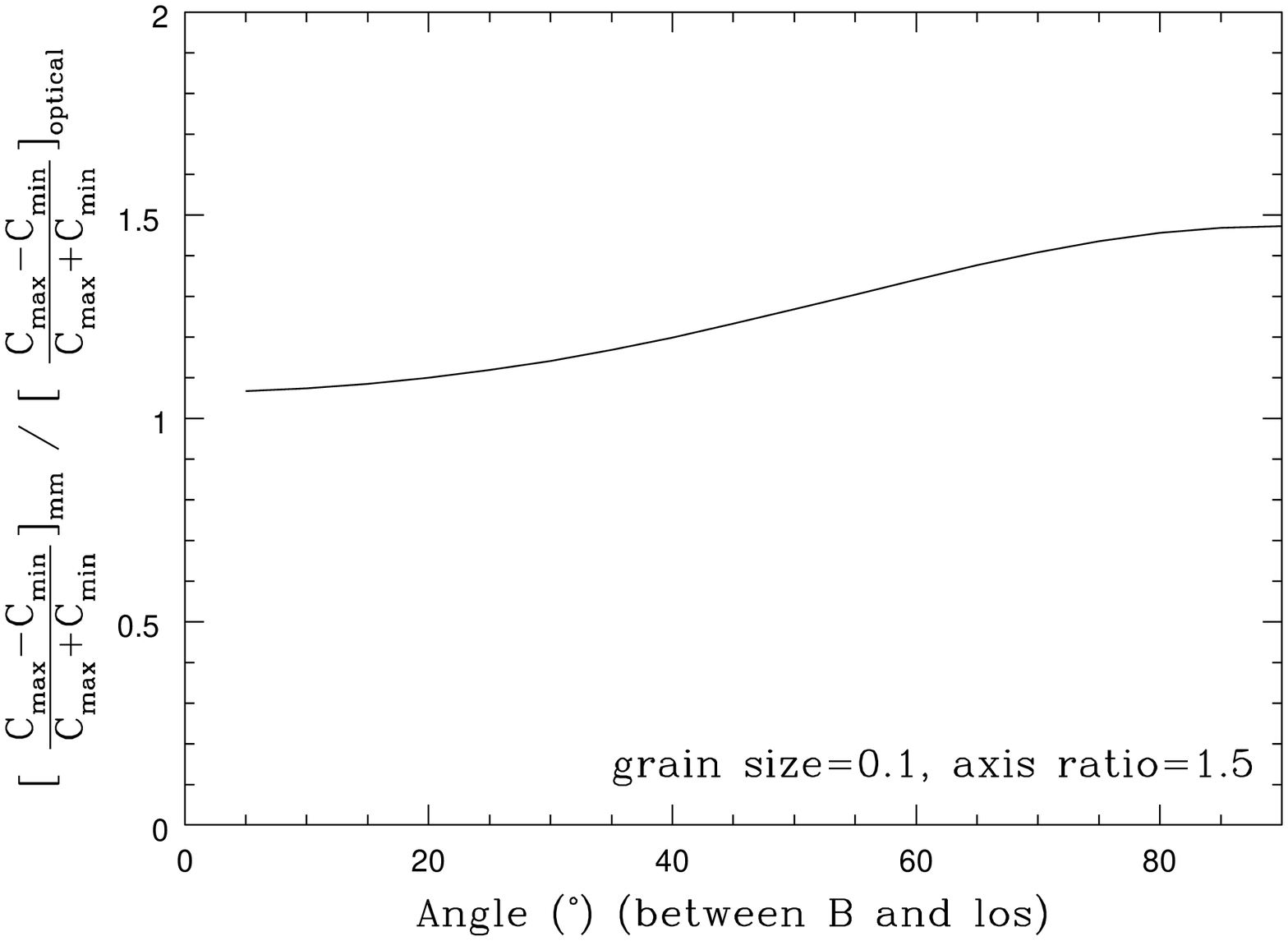}  
\caption{ 
   The ratio of $P_{em,mm}/P_{em,optical}$.
  {\it Left}: The polarization ratio vs.~the axis ratio ($a_2/a_1$) 
   of aligned oblate spheroidal grains.
   The polarization ratio shows only a weak dependence on the axis ratio.
   We assume that the grain are perfectly aligned with their long axes perpendicular to magnetic field,
the grain size is 0.1$\mu$m, magnetic field is perpendicular
   to the line-of-sight, $\lambda_{optical}=0.5\mu$m,
   and $\lambda_{mm}=1000\mu$m.
  {\it Right}: The polarization ratio vs.~the angle between magnetic field and 
   the plane of the sky.
   We assume that the grain size is 0.1$\mu$m, the grain axis ratio is 1.5,
   $\lambda_{optical}=0.5\mu$m,
   and $\lambda_{mm}=1000\mu$m.
}
\label{fig:ratios}
\end{figure*}

\subsection{Map of polarized dust emission from starlight polarization}

In general, it is advantageous to use all possible sources of information about 
foregrounds in order to improve their removal.  
In this section we discuss how the starlight polarization maps can be used 
to construct the maps of dust polarized emission. 
In principle, we can construct 
a polarized dust emission map at {\it mm} wavelengths ($I_{pol,mm}(l,b)$) from 
a dust total emission map ($I_{mm}(l,b)$) 
and a degree-of-polarization map ($P_{em,mm}(l,b)$) at {\it mm} wavelengths:
\begin{equation}
  I_{pol,mm}(l,b)= P_{em,mm}(l,b)~ I_{mm}(l,b),  \label{eq:IIP}
\label{eq_Ipol_mm}
\end{equation}
where $(l,b)$ denotes the Galactic coordinate.
However, neither $I_{pol,mm}(l,b)$ nor $I_{mm}(l,b)$ is directly
available.
Therefore, we need indirect methods to get 
$I_{pol,mm}(l,b)$ and $I_{mm}(l,b)$.

Obtaining a dust total emission map ($I_{mm}(l,b)$) is relatively easy
because dust total emission maps at FIR wavelengths are already available
{}from the {\it IRAS} and {\it COBE/DIRBE} observations.
Using the relation
\begin{equation}
     I_{mm}(l,b) = I_{100\mu m}(l,b)  ( 1mm/100\mu m )^{-\beta}, \label{eq:mm_100}
\end{equation}
where $1 \lesssim \beta \lesssim 2$,
one can easily obtain an emission map at {\it mm} wavelengths ($I_{mm}$) 
{}from the maps at 100 $\mu m$ or 240 $\mu m$.
However, more sophisticated model dust emission maps at {\it mm} wavelengths
already exist.
For example,
Finkbeiner et al. (1999) presented predicted full-sky maps of
microwave emission from the diffuse interstellar dust using
FIR emissions maps generated by Schlegel et al. (1998).
In fact, the model dust emission map we analyzed in the previous section
(\S\ref{sect:dust_em}) is one of the maps presented in 
Finkbeiner et al. (1999). 
Therefore, we can assume that the thermal dust emission map 
($I_{mm}(l,b)$) is already available.

Obtaining a degree-of-polarization map at {\it mm} wavelengths ($P_{em,mm}(l,b)$)
is relatively more complicated.
We can use measurements of starlight polarization at optical wavelengths
to get $P_{em,mm}(l,b)$. The basic idea is that
the degree of polarization by emission at mm ($P_{em,mm}$) is related to 
that at optical wavelengths ($P_{em,optical}$), which in turn is related to
the degree of polarization by absorption at optical wavelengths
($P_{abs,optical}$):
\be
  P_{abs,optical} \rightarrow P_{em,optical} \rightarrow P_{em,mm}.
\ee
 We describe the relations in detail below.

When the optical depth is small, we have the following relation (see, for
example, Hildebrand et al. 2000):
\begin{equation}
  P_{em,opt} \approx -P_{abs,opt}/\tau,   \label{eq:emabs}
\end{equation}
where $P_{em,opt}$ is the degree of polarization by emission and
$\tau$ is the optical depth (at optical wavelengths).
We obtain polarization by emission at {\it mm} wavelengths ($P_{em,mm}$) 
using the relation
\begin{equation}
  P_{em,mm}= P_{em,opt}
        \left[ \frac{ C_{max}-C_{min} }{ C_{max}+C_{min} } \right]_{mm}
      / \left[ \frac{ C_{max}-C_{min} }{ C_{max}+C_{min} } \right]_{opt},
  \label{eq_pem_mm}
\end{equation}
where $C$'s are cross sections (of grains as projected on the sky)
that depend on the geometrical shape (see, for example, the discussion in 
Hildebrand et al. 1999; see also Draine \& Lee 1984)
and dielectric function $\epsilon=\epsilon_1+i\epsilon_2$
(see Draine 1985) of 
grains.

For {\it mm} wavelengths, it is easy to calculate the ratio in Eq.~(\ref{eq_pem_mm})
because the wavelength $\lambda$ is much greater than the grain size $a$ 
(i.e.~$\lambda \gg 2\pi a$).
In this case, 
if grains are oblate spheroids with $a_1<a_2=a_3$
and 
short axes ($a_1$) of grains are perfectly aligned in the
plane of the sky, we have
\begin{equation}
    C_j=\frac{ 2\pi V }{ \lambda } 
        \frac{ \epsilon_2(\lambda ) }
        { \left( L_j \left[ \epsilon_1(\lambda )-1 \right] +1 \right)^2 
               + \left[ L_j \epsilon_2(\lambda ) \right]^2  },
\label{eq:c_j}
\end{equation}
where $L$ values are defined by
\bea
   L_1 &=& [(1+f^2)/f^2][1-(1/f)\arctan f], \nonumber \\
   L_2 &=& L_3=(1-L_1)/2, \nonumber \\
   f^2 &=& (a_2/a_1)^2-1
\eea
(see, for example, Hildebrand et al.~1999).

However, for optical wavelengths, the condition $\lambda \gg 2\pi a$ is {\it not}
always valid and, therefore, the expression in Eq.~(\ref{eq:c_j}) returns 
only approximate values.
For accurate evaluation of the cross sections, one should use numerical methods.
Fortunately, several numerical codes are publicly available for such 
calculations (for example, {\bf DDSCAT} package by Draine \& Flatau 1994, 2008; 
{\bf ampld.lp.f} by Mishchenko 2000).
We use {\bf ampld.lp.f} to calculate the ratio in Eq.~(\ref{eq_pem_mm}).
We assume that the grains are oblate spheroids, grain size is $0.1\mu$m, 
$\lambda_{optical}=0.5\mu$m, and $\lambda_{mm}=1000\mu$m.
Left panel of Fig.~\ref{fig:ratios} shows that the ratio is around $1.5$
when magnetic field is perpendicular to the line-of-sight. It also shows that
the ratio of 
$P_{em,mm}/P_{em,optical}$ is almost independent of the grain axis ratio.

In this subsection, we described a simple way to
obtain a polarized map at {\it mm} wavelengths.
However, actual implementation of the method can be more complicated
due to the following reasons.
First, we used an assumption that the grains that produce optical
absorption produce also microwave emission. But, this is not true
in general (see Whittet et al. 2008). 
Second, the expressions in Eqs.~(\ref{eq_pem_mm}) and (\ref{eq:c_j})
are valid when magnetic field direction is fixed and perpendicular 
to the line-of-sight and
all grains are perfectly aligned with the magnetic field.
If this is not the case, Eq.~(\ref{eq_pem_mm}) will become
$P_{em,mm} \propto P_{em,optical}$ with the constant of proportionality
that depends on magnetic field structure and the degree of
grain alignment. The effect of partial alignment is expected to be less
important\footnote{
 Grain alignment theory based on radiative torques 
 (Dolginov \& Mytrophanov 1976, Draine \& Weingartner 1996, 1997; 
 Weingartner \& Draine 2003; Lazarian \& Hoang 2007ab; Hoang \& Lazarian 2008, 2009ab; 
 see also Lazarian 2007, 2009 for reviews) predicts that
 grains starting with a particular size, which is $\sim 5\times 10^{-6}$~cm for the 
 typical interstellar radiation field, get aligned.
 If the grain size distribution varies from one place to another,
 this fraction of aligned grains will also vary.
 However, for the diffuse ISM, the grain distribution does not vary much
 (Weingartner \& Draine 2001). Therefore, the effect of variations of the 
 degree of grain alignment due the variations of the grain size is of secondary importance. 
 Much more important is the effect of incomplete alignment arising for some grains due to 
 their shape and the direction of the radiation and magnetic field predicted in Lazarian \& Hoang 2007a. 
 The consequences of this effect for polarization maps require more studies (see Padoan et al. 2009).}. 


The effect of non-perpendicular magnetic field can be potentially important.
We perform a numerical calculation using {\bf ampld.lp.f} to evaluate
the effect.
We assume that the grains are oblate spheroids, grain size is $0.1\mu$m, 
$\lambda_{optical}=0.5\mu$m, and $\lambda_{mm}=1000\mu$m.
Right panel of Fig.~\ref{fig:ratios} shows that the polarization ratio
drops from  $\sim 1.5$ to $\sim 1.1$ when the angle (between magnetic field
and the plane of the sky) changes from $90^\circ$ to $\sim 5^\circ$.
Therefore, the effect is not very strong and can be potentially corrected for\footnote{
   We can make use of Right panel of Fig.~\ref{fig:ratios} reversely.
   In the future, when we can accurately measure
   polarized emission from thermal dust in FIR or $mm$ wavelengths,
   we can obtain the values of $[(C_{max}-C_{min})/(C_{max}+C_{min})]_{mm}$.
   This result combined with the values of 
   $[(C_{max}-C_{min})/(C_{max}+C_{min})]_{optical}$ 
   in optical wavelengths can be used to 
   find average angle between magnetic field and the plane of the sky.
   That is, when we know the ratio $[...]_{mm}/[...]_{optical}$,
   we can use Right panel of Fig.~\ref{fig:ratios} to find
   the angle between magnetic field and the plane of the sky.}.

\subsection{Angular spectrum of polarized emission from thermal dust}
After we have constructed a map of the polarized emission from thermal dust, 
we can obtain the angular spectrum.
However, if we are interested in only the shape of the angular spectrum,
we do not need to construct the polarized thermal dust emission map.
We can get the shape of the the angular spectrum
directly from the starlight polarization map $P_{abs,optical}(l,b)$.

Eq.~(\ref{eq:IIP}) tells us that $I_{pol,mm}$ is given by  
$P_{em,mm}$ times $I_{mm}$. 
{}From Eqs. (\ref{eq:emabs}) and (\ref{eq_pem_mm}), we have
\bea
   I_{pol,mm}& = & P_{em,mm}~ I_{mm} \propto P_{em,opt}~ I_{mm}  \nonumber \\
   & \approx & (P_{abs,opt}/\tau)~ I_{mm} \nonumber  \\
   & \propto & P_{abs,opt}.  \label{eq:28}
\eea
Here we use the fact $\tau \propto I_{mm}$.
Note that the constant of proportionality does not affects the shape of the
angular spectrum if grain properties do not vary much in halo.
Therefore, as to the power spectrum $c_l$ of $I_{pol,mm}$, we can use
that of $P_{abs,optical}$:
\be
    C_l \mbox{~of $I_{pol,mm}$} \propto C_l \mbox{~of $P_{abs,opt}$}.
   \label{eq:29}
\ee
Once we know the angular spectrum of $P_{abs,optical}$,
we can estimate the angular spectrum of $I_{pol,mm}$.

Angular spectrum of starlight polarization, $P_{abs,opt}$(l,b), is already available.
Fosalba et al.~(2002) 
obtained $C_l \sim l^{-1.5}$ for starlight polarization.
The stars used for the calculation are at different distances from
the observer and most of the stars are nearby stars.
The sampled stars are mostly in the Galactic disk.
CL02 reproduced the observed angular spectrum numerically
using mixture of stars with a realistic distance distribution.
CL02 also showed
that the slope becomes shallower when only stars with a large fixed distance
are used for the calculation. 
Therefore, it is clear that distance, 
or dust column density, 
to the stars is an important factor
that determines the slope. 
We expect that, if we consider only the nearby stars with a fixed distance,
the slope will be steeper.
This means that, if we consider stars in the Galactic halo,
the slope will be steeper.

The method described above requires measurements of polarization from
many distant stars in the Galactic halo.
Unfortunately, the number of stars outside the Galactic disk that can be
used for this purpose are no more than a few thousands (Heiles 2000; see also discussions in
Page et al. 2007; Dunkley et al. 2008).
When more observations are available, accurate estimation
of $I_{pol,mm}(l,b)$ (and $C_l$ of $P_{abs,optical}$) will be possible.


\subsection{Model calculations for starlight polarization} \label{subsect_starpol}

We expect the fluctuations of the starlight polarization to arise 
primary from the fluctuations of magnetic fields\footnote{In addition, 
fluctuations arising from the variations of the degree of grain alignment (see Lazarian 2007) 
are expected.}. 
The latter are expected to have the spectral index close to the Kolmogorov one (see \S 3). 

As we discussed in the previous subsection, the angular spectrum, $C_l$,
of $P_{abs,optical}$ for the Galactic halo will be different from the observed $l^{-1.5}$ 
spectrum for mixture of stars with different distances in the Galactic disk.
However, it is not clear exactly how the former is different from the latter.
To deal with this problem we use numerical simulations again.
We first generate two sets of magnetic field on a two-dimensional
plane ($8192 \times 8192$ grid points), 
using Kolmogorov three-dimensional spectra\footnote{
   Consider a 3-dimensional magnetic field with a 3D spectrum
   $E_{3D}(k_x,k_y,k_z)$ ($\propto k^{-m}$ with $m=11/3$ for Kolmogorov turbulence). 
   The spectrum of the magnetic field on a two-dimensional
   sub-plane (e.g.~$z=0$ plane) is 
   $E_{z=0~plane}(k_x,k_y)\propto 
     \int_{-\infty}^{\infty} dk_z~E_{3D}(k_x,k_y,k_z)$, which
   we use to generate two sets of magnetic field
   on a two-dimensional plane. 
   Note that, although this spectrum does not follow a power law
       near the outer scale of turbulence, it is close to 
       $k^{-m+1}$ for large values of $k$.
}.
Since we need $P_{abs,optical}$ for stars well above the Galactic disk, 
we assume the distance to stars is fixed in each model.
We consider 3 models:
\begin{enumerate}
\item {\bf Case 1}, Nearby stars in a homogeneous turbulent medium: 
       We generate three (i.e.~{\it x,y}, and {\it z}) 
       components of magnetic field on a two-dimensional
       plane ($8192 \times 8192$ grid points representing 
       $400 pc \times 400 pc$), 
       using the following Kolmogorov three-dimensional spectrum:
       $E_{3D}(k)\propto k^{-11/3}$ if $k>k_0$,
       where $k_0 \sim 1/100~pc$. 
     (The outer scale of turbulence is 100pc.)
      We assume the volume density of dust is homogeneous. 
      All stars are at a fixed distance of $100pc$ from the observer. 
\item {\bf Case 2}, Distant stars in a homogeneous turbulent medium:
       We generate three (i.e.~{\it x,y}, and {\it z}) 
       components of magnetic field on a two-dimensional
       plane ($8192 \times 8192$ grid points representing 
       $4~kpc \times 4~kpc$), 
       using the following Kolmogorov three-dimensional spectrum:
       $E_{3D}(k)\propto k^{-11/3}$ if $k>k_0$,
       where $k_0 \sim 1/100~pc$. 
      Other setups are the same as those of Case 1, 
      but the distance to the stars is $2kpc$.
\item {\bf Case 3}, Stars in a stratified medium: 
      We use the magnetic field generated in Case 1.
      The volume density of dust shows a $sech^2(z)$ decrease: 
      $\rho(r)=4\rho_0 /[ \exp(r/r_0)+\exp(-r/r_0) ]^2$.
      We assume spherical geometry and $r_0 = 100pc$. 
      The stars are at $r=200pc$ from the observer.
      The outer scale of turbulence is 100pc.
\end{enumerate}

We assume that dust grains are oblate spheroids.
In the presence of a magnetic field, some grains (especially large grains)
are aligned with the magnetic field (see Lazarian 2007 for a review). 
Therefore, cross sections parallel
to and perpendicular to the magnetic field are different.
We assume that parallel cross section is $\sim$30\% smaller than the 
perpendicular one.
We use the following equations to follow changes of Stokes parameters
along the path:
\bea
 I^{-1} dI/ds & = & -\delta + \Delta \sigma Q/I, \\
 d(Q/I)/ds & = & \Delta \sigma - \Delta \sigma (Q/I)^2, \\
 d(U/I)/ds & = & -\Delta \sigma (Q/I)(U/I)
\eea
(see Martin 1974 for original equations; see also Dolginov, Gnedin, \& Silantev 1996),
where $\delta = (\sigma_{1} + \sigma_{2})$,
$\Delta \sigma = (\sigma_{1} - \sigma_{2})$, and
\bea
2\sigma_{1} &=& \sigma_{\perp}, \\
2\sigma_{2} &=& \sigma_{\perp} -(\sigma_{\perp}-\sigma_{\|}) \cos\gamma
\eea
(Lee \& Draine 1985).
Here $\sigma_{\perp}$ and $\sigma_{\|}$ are the extinction coefficients 
and $\gamma$ is the angle
between the magnetic field and the plane of the sky. 
After we get the final values of Stokes parameters, we calculate
the degree of polarization ($\sqrt{Q^2+U^2}/I$) and, then, the second-order 
angular structure function of the degree of polarization.

\begin{figure*}[h!t]     
\includegraphics[width=0.48\textwidth]{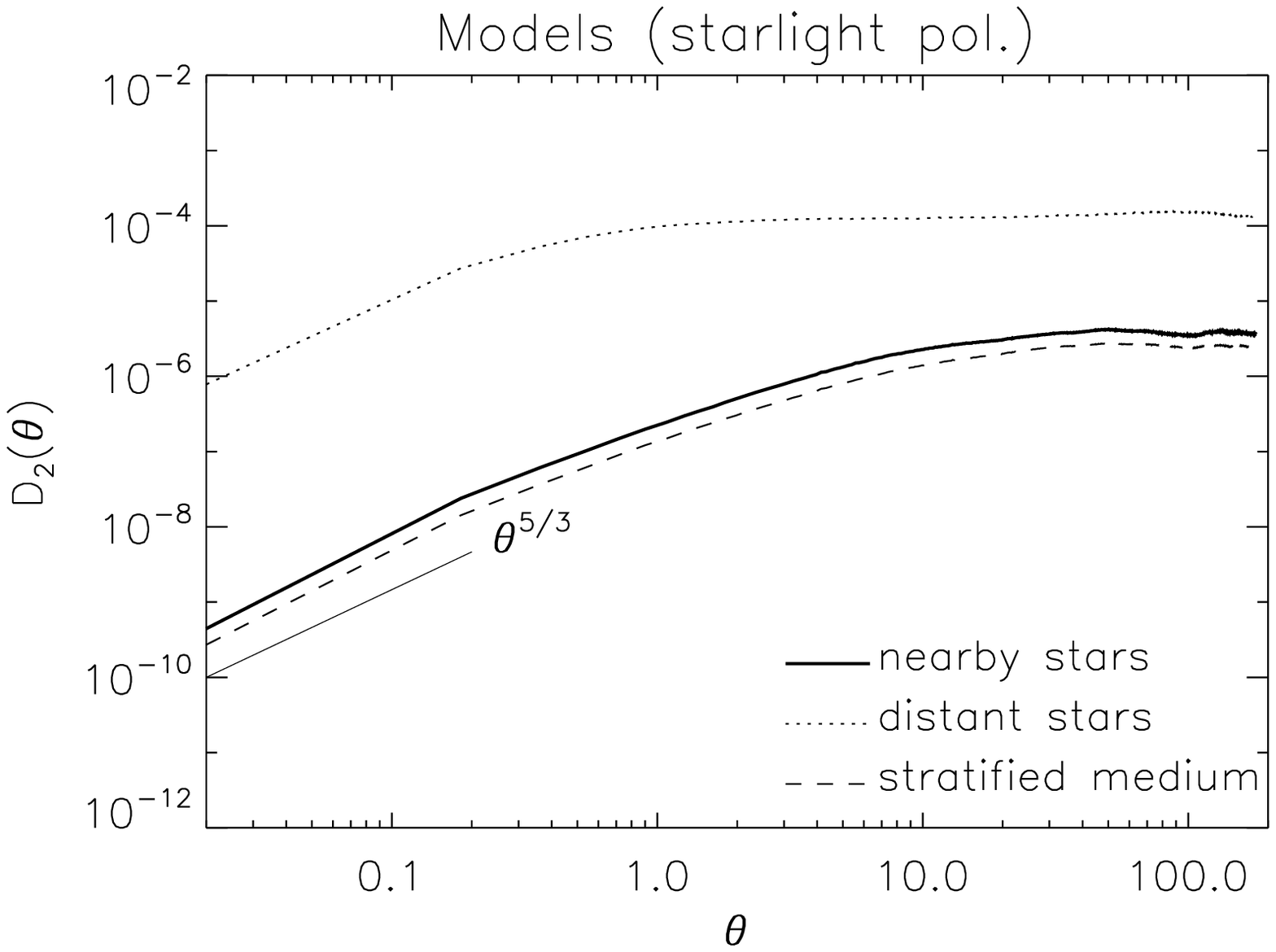}   
\includegraphics[width=0.48\textwidth]{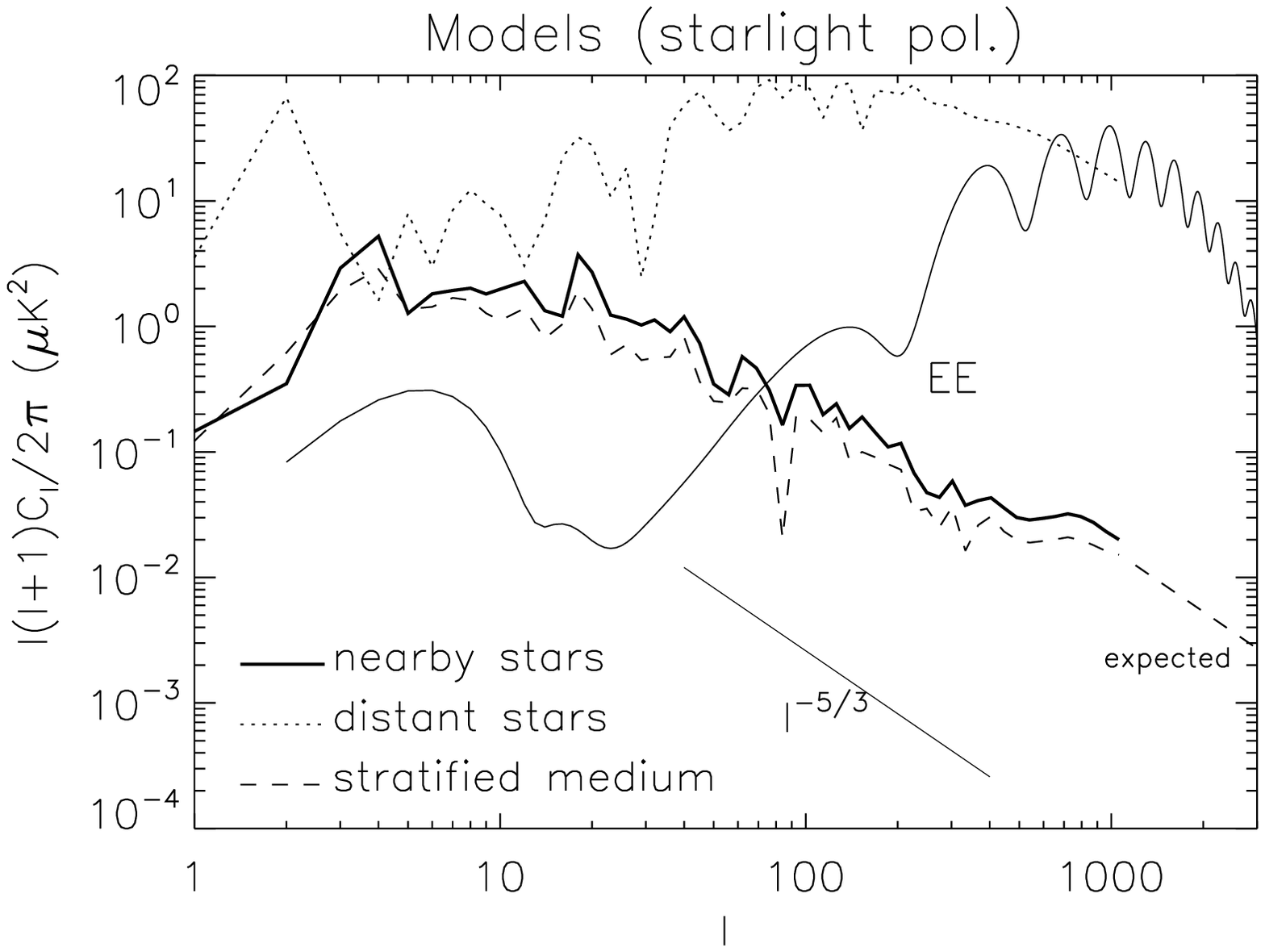}   
\caption{ 
     Model calculations for starlight polarization.
     Figures show the second-order angular structure functions (left) 
     and the angular power spectra (right)
     for degree of polarization (right).
     {\it Left}: The second-order structure functions from model calculations.
     The structure function for
     nearby stars (thick solid line; $d=100$pc; Case 1 in the text) 
     has a slope close to Kolmogorov one for $\theta \lesssim 0.2^\circ$ 
     and slightly
     shallower spectrum for $0.2^\circ \lesssim \theta \lesssim 10^\circ$.
     The case for density stratification 
    (dashed line; scale height $= 100$pc; Case 3 in the text)
     shows also a similar slope.
     However, the case for far-away stars 
    (dotted line; $d=1$kpc; Case 2 in the text)
     shows a substantially shallower slope for $\theta \gtrsim 0.1^\circ$.
     {\it Right}: 
        The angular power spectra from model calculations.
        We take the shape of the spectrum from Section \S\ref{subsect_starpol} and
        the amplitude from Page at al.~(2007) result.
        We assume that the observed frequency is $94 GHz$.
        Nearby stars (thick solid line) or 
        stars in stratified medium (dashed line) 
        show a slope flatter than
        the Kolmogorov when $l\lesssim 1000$. 
        The second-order structure function on the left panel implies that
        $l(l+1)C_l \propto l^{-5/3}$ for $l>1000$.
        The straight dashed line for $l>1000$ is obtained this way.
     Note that $\theta=0.1^\circ$ corresponds to 
      $l\sim 180^\circ/\theta^\circ \sim 1800$.
     We also show the polarized CMB `EE' spectrum.
 }
\label{fig:starpol}
\end{figure*}

We show the result in Fig.~\ref{fig:starpol}.
When all the stars are at the distance of 100$pc$ (Case 1), the spectrum is consistent with the Kolmogorov
one for small $\theta$.
The result for the stratified medium (Case 3) also shows a spectrum compatible with
the Kolmogorov one for small $\theta$.
When stars are far away (Case 2), the qualitative behavior is similar.
However, if we measure average slope between $\theta = 0.2^\circ$ and $20^\circ$, 
the result is different:
the slope for Case 2 is substantially shallower.
Note that $\theta = 0.2^\circ$ and $20^\circ$
correspond to $l=1000$ and $10$, respectively.
This means that,
 when we have distant stars only, 
the angular spectrum will be shallower than the Kolmogorov one. 
When we have a mixture of distant and nearby stars, we will have
an angular spectrum that is steeper than the case of distant stars
but shallower than the case of nearby stars, which implies the spectrum is
shallower than Kolmogorov one.
Therefore, it is not surprising that Fosalba et al.~(2002) obtained
a shallow spectrum of $\sim C_l\propto l^{-1.5}$ 
for a mixture of nearby and distant stars mostly in the Galactic disk.
Flattening of spectrum (i.e. $C_l \propto l^{-\alpha}$ with
$\alpha \approx 1.3$ $\sim$ $1.4$) for polarized FIR dust thermal emission is also
observed in Prunet et al.~(1998; see also Prunet \& Lazarian 1999).

Note that the spectrum of the emission polarization 
for very large values of multipole $l$ is expected to be 
steeper than the
spectrum of starlight polarization measured at small values of $l$ 
For instance, from our model calculations predict that we should see the Kolmogorov spectrum of polarized emission, 
rather than $\sim C_l\propto l^{-1.5}$. 
This difference must be taken into account if starlight polarization is used 
to filter the polarized microwave emission arising from dust. In fact, to do filtering one 
should either make a model of the magnetic field distribution based on the extended samples 
of stars throughout the Galactic volume or use only distant stars to get the spectrum of 
polarization similar to that expected at the microwave range.

There could be systematic errors in transforming   
{}from the starlight polarization spectrum to that of emission.
One of such errors may arise from the variation of the direction of the mean magnetic field direction 
and the line of sight. 
We expect that, while the starlight polarization spectrum is relatively insensitive
to the mean magnetic field direction, the emission spectrum shows a stronger dependence
on it. 
Our preliminary calculations show that such a systematic error is small.
We will pursue this possibility in the future.

\subsection{Comparison with the CMB polarization}
On the right panel of Fig.~\ref{fig:starpol}, we plot angular power spectrum
of star light polarization.
As we mentioned earlier, the angular spectrum of the degree of starlight polarization
should be similar to that of polarized thermal dust emission 
(see Eqs.~[\ref{eq:28}] and [\ref{eq:29}]).

To obtain angular spectra, we use a Gauss-Legendre quadrature integration
method as described in Szapudi et al. (2001).
To be specific, we first generate magnetic fields from the 3 models we considered 
in the previous subsection. 
Then, we calculate angular correlation functions, $K(\cos\theta)$.
Finally, we obtain the angular spectra using Eq.~(\ref{eq:k2cl}).
Since $C_l$ obtained in this way is very noisy, we plot $C_l$ averaged over the multipole range 
$(l/1.09, 1.09l)$.
We do not show $C_l$ for $l>1000$ because it is too noisy even with the averaging process.
        The second-order structure function on the left panel of Fig.~\ref{fig:starpol} implies that
        $l(l+1)C_l \propto l^{-5/3}$ for $l>1000$.
        The straight dashed line for $l>1000$ reflects this implication.
        
We normalize the spectra using the condition
$\sum_{2}^{10} (l+1)C_l/2\pi = 3 (\mu K^2)$ for the acse of stars in the stratified medium.
This normalization is based on the values given in Page et al.~(2007; their Eq.~[25])\footnote{
   The purpose of this normalization is to match roughly our spectrum and that of Page et al.~(2007)
   on large scales.
   Eq.~(25) of Page et al.~(2007) reads $(l+1)C_l^{dust}/2\pi = 1.0(\nu/65 GHz)^3 l^{-1.6} 
       \sim 3 l^{-1.6} ~(\mu K^2)$ for $\nu=94$ GHz.
   When summed from $l=2$ to $l=10$, this gives $\sim 2.6 (\mu K^2)$.
   We adopt $\sum_{2}^{10} (l+1)C_l/2\pi = 3 (\mu K^2)$ for the case of stars in the stratified medium.
}.
We assume that the observed band is W-band ($\nu = 94 GHz$).

The plot shows that the slopes for nearby stars (thick solid line) 
and stars in the stratified medium are shallower than that of 
the Kolmogorov spectrum
for $l<1000$. 
This result is consistent with that obtained with the angular structure function.
Note that the case of distant stars has much flatter spectrum for $l<1000$.
We believe that our toy model for the stratified medium (Case 3) 
better represents the actual situation for polarized emission from thermal 
dust in the Galactic halo. 
Therefore, 
we expect that the polarized thermal emission from thermal dust
in high-latitude Galactic halo has a spectrum slightly shallower than the Kolmogorov spectrum
for $l<1000$.
Our calculation do not tell us about the slopes for $l>1000$.
However, judging from the behavior of the angular structure function 
for $\theta \lesssim 0.1^\circ$,
we expect that $C_l \propto l^{-11/3}$ for $l>1000$
(see the straight dashed line for $l>1000$ on the right panel of Fig.~\ref{fig:starpol};
see also discussions in \S\ref{sect:6_4}).

We also show the polarized CMB `EE' spectrum in the Figure
(data from CMBFAST online tool at http://lambda.gsfc.nasa.gov/).
The Figure shows that the EE spectrum dominates polarized thermal dust emission from high-latitude
Galactic halo for
$l\gtrsim 100$. The EE spectrum is expected to be sub-dominant when $l>5000$.

\section{Spatial filtering of foregrounds}  \label{sect_filter}

A unifying theme of this study and that in CL02 is that the principal source 
of the foreground fluctuations is related to the MHD turbulence in the Galactic 
interstellar medium. 
Our analysis of the available observational data in both publications supports this conclusion. 
In this section we discuss how this insight into the origin of the foreground fluctuations 
can be used to remove the foregrounds.  

\subsection{Statistical properties of foregrounds} 

Removal of Galactic foregrounds has always been a major concern for CMB
studies. The challenge is only going to increase substantially now, when CMB
polarization studies are attempted.

The knowledge that foregrounds are not an arbitrary noise, but have well
defined statistical properties in terms of their {\it spatial power spectra}
is an important additional information that can be utilized to evaluate and
eventually eliminate the foreground contribution. 

Utilizing the information about underlying turbulence power spectrum is not
straightforward, however. Our study shows that the observed power spectrum
may depend on geometry of the emitting volume. Therefore, the detailed
modeling of the foreground fluctuations should involve accounting for the
geometry of the emitting volume.

The latter point stresses the synergy of Galactic foreground and CMB
studies. Indeed, our fitting of the power spectra 
in Fig.~\ref{fig:model} shows that on the
basis of its variations we may distinguish between different models of the
emitting turbulent volume. As soon as this achieved, one can {\it predict},
for instance the level of fluctuations that are expected from the foreground at the
scales smaller than those studied. 

While the previous statements are true in general terms, a number of special cases in which 
simpler analysis is applicable are available. For instance, a simplification that is expected at
higher resolutions that are currently available, is that at sufficiently
small scales the statistics should get independent of the large-scale
distributions of the emitting matter. Moreover, simple power laws are expected and observed 
(see Figure~\ref{fig:sfetc-has}) for the foregrounds at high Galactic latitudes. 
Therefore modeling of the distribution of the Galactic emission and the filtering 
of it using the approach in \S 3 are not necessarily interlocked problems.

\subsection{Examples of filtering}

We remind the reader that the approach in \S 3 requires first separating a particular component of 
a foreground over a range of scales\footnote{If we know the foreground at a single spatial scale, 
one can use the a priori knowledge of the expected spatial  scaling of the foreground spectrum. 
For instance, if underlying fluctuations are related to magnetic field they are likely to have 
the spectrum close to the Kolmogorov one.}, e.g. using the traditional technique of frequency templates, 
and then extending the spatial scaling of the foreground's $C_l^F$ to higher $l$.
Consider a few examples of utilizing this approach.

For instance, high resolution measurements of the South Pole Telescope (SPT) (see Lueker et al. 2009) 
and the Atacama Cosmology Telescope (ACT) (see Fowler et al. 2010) will provide measurements at high $l$, 
but will have limited frequency coverage to remove the foregrounds. Therefore, the spatial 
extrapolation of $C_l^F$ obtained with other low resolution experiments, 
which, however, provide good frequency coverage required for the $C_l^F$ identification may be advantageous. 

Prior to the release of the {\it Planck} data, the extension of $C_l^F$ obtained, for instance, 
with the WMAP to higher $l$ may be useful for the foreground filtering for the suborbital missions.    
After the release of the {\it Planck} data, the studies of the advocated approach can still be useful. 
Consider, for instance, Fig.~\ref{fig:starpol}. 
If {\it Planck} measures the spatial spectrum up to
$l=2000$, then for a higher resolution balloon mission one can evaluate the
level of foreground contamination by extrapolating the expected foreground
spectrum. Moreover, the study of the B-modes would require new experiments with higher sensitivity 
and the extrapolation procedure can be again useful. 

\subsection{Demonstration of filtering technique 1} \label{sect_demo}

\begin{figure*}[h!t]
\includegraphics[width=0.85\textwidth]{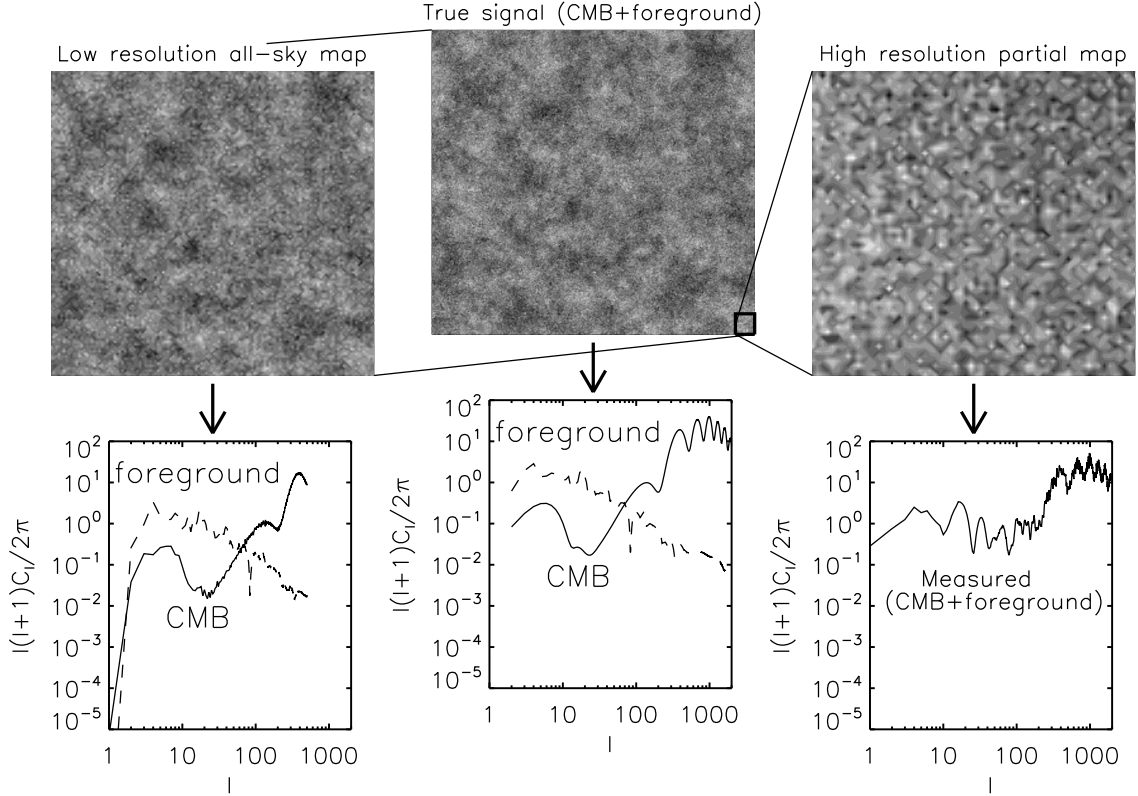}  
\caption{ Explanation of our filtering process.
    We use a Cartesian coordinate system for simplicity and assume there is only
    one foreground.
   {\it Left panels}:
    Low angular resolution all-sky map ($1024 \times 1024$ pixels; $\Delta \theta \sim 21^\prime$).
    We generate the low resolution map by applying a circular top-hat filter
    on the data shown in the middle panels. 
    The radius of the filter is 4 grid units, which is equivalent to $\sim 10.5^\prime$
    The size of the resulting data file is $1024 \times 1024$, the angular
    resolution of which corresponds to $\Delta \theta \sim 21^\prime$.
    We obtain the spectra in the lower-middle panel by direct Fourier transform. 
   {\it Middle panels}: 
    All-sky map of the CMB and the foreground signals and their 
    angular spectra.
    We generate the map in the upper-middle panel using the spectra shown in lower-middle panel.
    The actual resolution of the map we use for the calculation is 
    $8192\times 8192$. 
   {\it Right panels}: 
     High angular resolution partial-sky map ($128 \times 128$ pixels; $\Delta \theta \sim 5.2^\prime$).
     This data file is from the low-right corner of the original 
     $8192 \times 8192$ data.
     We obtain this map by skipping every other point in each
     direction, so that the angular resolution 
     of this data ($\Delta \theta \sim 5.2^\prime$)
     is twice worse than the original  $8192 \times 8192$ data, but
     4 times better than the low-resolution data in middle panels.
     On the original data file (i.e. on the $8192 \times 8192$ data), 
     the region spans $256 \times 256$ grid points.
     Therefore, the size of the partial-sky map is $128 \times 128$.
     The angular size of map in the upper-right panel is
     $11^\circ \times 11^\circ$. 
    The maps shown in this figure are degraded to reduce file size.
}
\label{fig:expl}
\end{figure*}
\begin{figure}[h!t]
\includegraphics[width=0.45\textwidth]{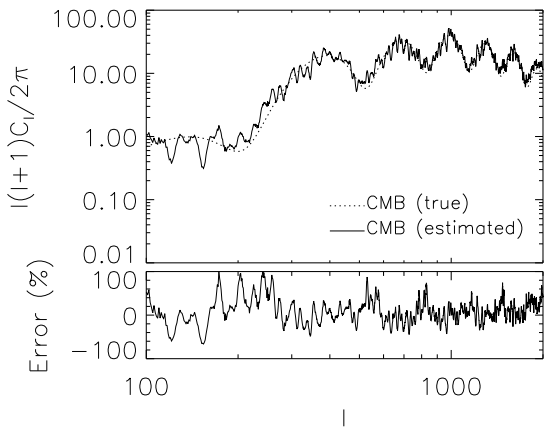}  
\caption{Angular spectrum of CMB obtained from the filtering process.
   {\it Top:} The solid line is the spectrum obtained from the filtering process and
   the dotted line is the original CMB spectrum.
   Note that, 
   since the high-resolution data span only $11^\circ \times 11^\circ$,
   the estimated angular spectrum for $l\lesssim 100$ is meaningless.
   {\it Bottom:} The solid line represents relative percentage error
   between the estimated CMB (solid curve) and the true CMB (dotted curve) spectra.
}
\label{fig:estimated}
\end{figure}

In this subsection, we demonstrate how the filtering process works.
For simplicity we use a Cartesian coordinate system.
For demonstration purposes, we consider the CMB EE spectrum (see the thin-solid line in Fig.~\ref{fig:starpol}b)
and Galactic dust foreground for the case of stratified medium (see the dashed line in Fig.~\ref{fig:starpol}b).

Suppose that low-resolution all-sky maps 
are already available for many frequency channels.
Let us assume the resolution of the maps is $\Delta \theta \sim 21^\prime$,
which is similar to the WMAP resolution.
Since there are many channels, one may use the usual filtering techniques
to separate the CMB and the foreground signals.
However, since we already know both the CMB and the foreground spectra in advance in this example,
we do not follow the usual filtering techniques.
Instead, we just assume that we already know the CMB and the foreground 
spectra for $l\lesssim 500$.
 
The `all-sky' map in the upper-left panel of Figure \ref{fig:expl}
represents this low-resolution map.
The map is defined on a grid of $1024 \times 1024$ points.
The angular resolution of the map is $\Delta \theta \sim 21^\prime$.
We generate the map from a much higher-resolution all-sky map (see upper-middle panel), 
which has a dimension of $8196 \times 8196$ in this example.
When we obtain the low-resolution all-sky map from the much higher resolution all-sky map, 
we apply a circular top-hat beam pattern with radius $\sim 10.5^\prime$ 
to mimic actual observation.
We calculate the spectra of the CMB and the foreground by
direct Fourier transform of the low-resolution map data.

Then, suppose that we perform a high-resolution balloon experiment that can cover 
only part of the sky. In such balloon experiments, frequency channels are
usually limited and removing foregrounds is a challenging task.

Here we show that, if we know the foreground spectrum, we
can easily obtain the CMB spectrum from the balloon data.
The upper-right panel  of Figure \ref{fig:expl} represents
the high-resolution balloon data.
The size of the data is $128 \times 128$.
The angular resolution is $\Delta \theta \sim 5.2^\prime$ and the map
covers $11^\circ \times 11^\circ$ in the sky.
We obtain the map by skipping every other point in each direction
on low-right corner of the original map with $8192\times 8192$ points.

Our goal is to obtain the angular spectrum of the CMB signal using
the balloon data.
We first make the balloon data periodic by proper reflections and translations
and obtain a periodic data on a grid of $4096 \times 4096$ points.
Then we multiply the data by a Gaussian profile of width $\sim 11^\circ$.
The center of the Gaussian profile should locate near the center of
the newly constructed $4096 \times 4096$ pixel data and coincide with the center of a $128\times 128$ pixel
original data.
Then, we
perform Fourier transform of the resulting data of size $4096 \times 4096$.
This way, we obtain an angular spectrum of the total 
(i.e.~CMB+foreground) fluctuations. 

Now, it is time to derive an angular spectrum of the foreground.
We first take the angular spectrum from the low-resolution map, which
is already available (see lower-left panel of Figure \ref{fig:expl}).
The dashed curve is the foreground spectrum.
The foreground spectrum is not defined for $l \gtrsim 500$.
In principle, we know the foreground spectrum when we know geometry and 
turbulence spectrum.
Here, we simply assume that the spectrum for $l\gtrsim 500$ is
Kolmogorov. 
This way, we can construct the foreground spectrum for all values of $l$.

The remaining task is just subtraction: when we subtract the foreground
spectrum from that of the total fluctuations, we can get the CMB spectrum.
The solid curve in Figure \ref{fig:estimated} is the CMB
spectrum obtained this way.
The solid curve show a very good agreements with the original CMB spectrum 
used for generating the original all-sky data on a grid of
$8192 \times 8192$ points (see middle panels).
The r.m.s. relative percentage error ($100\Delta C_l/C_l^{CMB}$) for $100< l < 2000$ is 24\%,
where $\Delta C_l$ is the difference between the estimated CMB spectrum ($C_l^{estimated}$; solid curve)
and the true CMB spectrum ($C_l^{CMB}$; dotted curve).

Note that the foreground spectrum depends on geometry of the
emitting regions and underlying turbulence spectrum. 
In earlier sections, we discussed geometry of light-emitting regions.
The turbulence spectrum may have some uncertainties (see Introduction).
Kolmogorov spectrum may be good for a number of cases.
However, it is possible that the spectrum deviated from Kolmogorov one
in some regions, such as molecular clouds.
However, such deviation will not affect our filtering process much
if the resolution of balloon data is not far beyond that of the low-resolution 
all-sky map.

\subsection{Demonstration of the filtering technique 2}  \label{sect_demo2}

\begin{figure*}[h!t]
\includegraphics[width=0.45\textwidth]{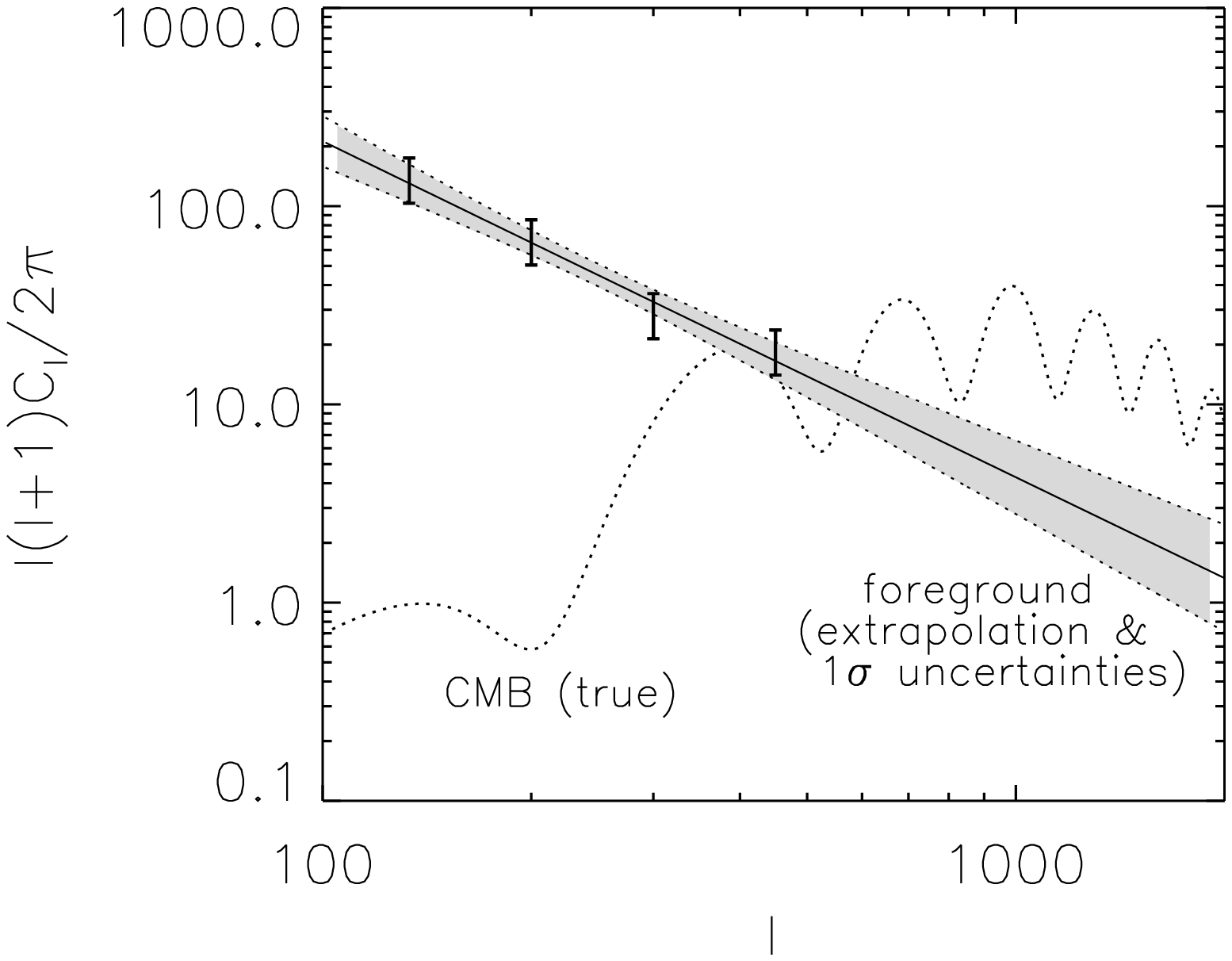}  
\includegraphics[width=0.45\textwidth]{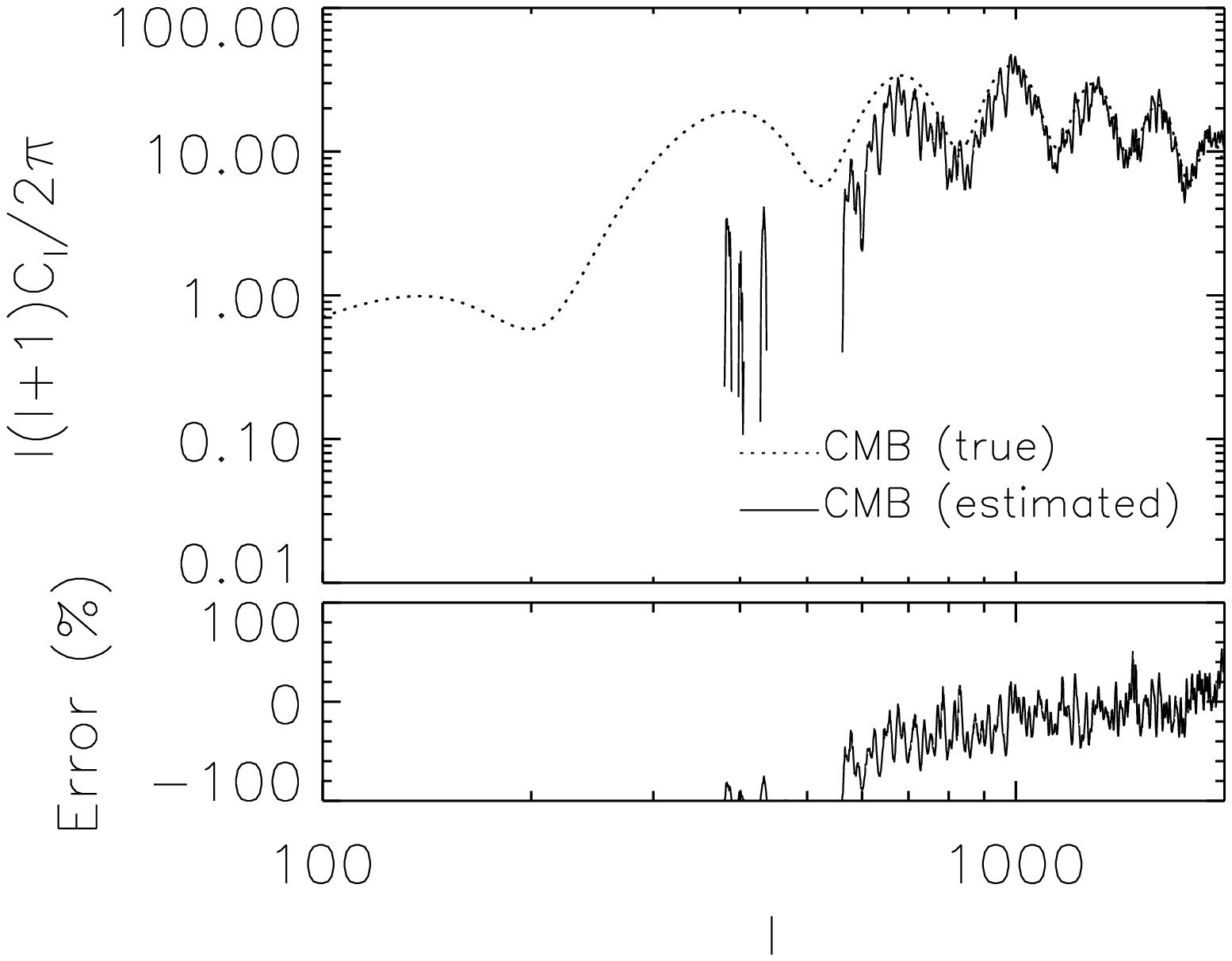}  
\caption{Second demonstration of the spatial filtering technique.
   For demonstration purposes, the foreground signal has been enhanced by 1000 times.
   {\it Left:} Most-likely foreground spectrum (solid line) and its $1\sigma$ uncertainties (shaded region).
   We obtain the spectrum and the uncertainties using the observed data.
   In this example, we mark the 4 observed point with error bars.
   See the text for details.
   {\it Right:} CMB spectrum obtained from the spatial filtering technique.
   It shows that the spatial filtering technique works for $l \gtrsim 550$, where
   the CMB and the foreground spectra have similar values. This result implies that the technique will work
   when the CMB spectrum is slightly larger than the foreground one.
}
\label{fig:error}
\end{figure*}

In the previous subsection, we demonstrated a reconstruction of CMB spectrum.
The result is very promising because the dust foreground model we used is not
far from a realistic one. Note that we normalized the amplitude of the dust foreground spectrum 
using
the value quoted in an earlier work (Page et al.~2007).

However, there are a couple of issues regarding to the previous demonstration.
First, in the previous demonstration, we assumed that we know the exact power-law index of 
the foreground spectrum.
In general, we may not know the exact power-law index of the foregrounds.
When this is the case, we need to extrapolate the foreground spectrum
to larger values of multipoles and 
figure out to what extent we can safely extrapolate.
Second, it is necessary to test how well our method works when the CMB and the foreground 
spectra are comparable.

In this subsection, we develop a method of deriving a most probable extrapolation of the foreground spectrum.
We also test the performance of our technique for
the case the CMB and the foreground 
spectra are comparable.
For this purpose, we scale up the amplitude of the foreground spectrum by a factor of 1000.\footnote{
   The motivation we take the factor of 1000 is that the CMB BB spectrum by weak gravitational lensing
   is about 100-1000 times smaller than the CMB EE spectrum for $l \lesssim 1000$ 
   (e.g.~Zaldarriaga \& Seljak 1998).   
   }
   
As in the previous subsection, we assume that multi-channel observations are available 
for $l \lesssim 500$. Therefore, we have foreground observation for $l \lesssim 500$.
However, unlike the previous subsection, we assume that observations are available 
only for selected values of multipole $l$. We mark 4 such $l$ values  
on Fig.~\ref{fig:error}. We use the same foreground spectrum as in the previous subsection
to generate the 4 observed points.
We also show the $1\sigma$ observation error bars for them, which are arbitrarily assigned in this example.

Using the 4 observed points for foreground, we find the most probable extrapolation for $l>500$.
It is tempting to use the linear least-square fit in a $\log$-$\log$ plane,
in which x-axis is a logarithm of $l$ and y-axis is that of $l(l+1)C_l/2\pi$.
But, it is not easy to estimate uncertainties in this case.

Therefore, we adopt a slightly different method.
We work on the $\log$-$\log$ plane. In this subsection, we use the following conventions: 
$x \equiv \log l$, $y\equiv \log [l(l+1)C_l/2\pi]$ and $(x_i, y_i)$ ($i=1,...,N$, where $N=4$ in this example) 
denotes an observed point.
For a given $x$, we want to find the most probable value and $1\sigma$ uncertainty of $y$.
In order to find them, we follow the following procedure.
First, we select an arbitrary $y$ and consider all possible lines that pass through
$(x,y)$. Let the slope of such a line be $a$ and the y-intercept be $b$.
Then, we calculate the following probability for the chosen $y$:
\be
     P(y)\equiv \exp\left[ -\sum_{i=1}^{N} \frac{(y_i - z_i)^2}{2\sigma_i^2} \right],
\ee
where $z_i=ax_i+b$ and $\sigma_i$ is the $1\sigma$ observation error for $i$.
Second, we repeat similar calculations for different values of $y$.
Third, we find the value of $y$ where $P_{tot}(y) \equiv \sum_{all lines} P(y)$ has the maximum.
Fourth, we find $1\sigma$ of the distribution $P_{tot}(y)$.
Fifth, we repeat similar calculations for different values of $x$.

We plot the result of this procedure in Fig.~\ref{fig:error}(a).
The solid line denotes the most probable values of `$y$' and 
the shaded region represents the $1\sigma$ uncertainty.
The $1\sigma$ uncertainty gets larger as $l$ gets larger when $l>500$.
Therefore, it may not be a good idea to extrapolate for $l$'s an order of magnitude larger than $500$
which is the maximum $l$ of the low-resolution multi-channel maps in our current example.
However, it seems to be OK to extrapolate for $l$'s a few times larger than $500$
in our current example.
We can reduce the $1\sigma$ uncertainty
by reducing the observation error bars, which may enable us to extrapolate further for larger $l$'s.

As in the previous subsection, we subtract the most-likely foreground spectrum 
{}from the observed (i.e. CMB+foreground) spectrum.
We plot the result in Fig.~\ref{fig:error}(b).

Fig.~\ref{fig:error}(b) shows that our technique recovers the 
CMB spectrum quite well for $l \gtrsim 550$.
Note that Fig.~\ref{fig:error}(a) tells us that the CMB and the foreground signals are 
almost same at $l\sim 550$.
This example implies that our technique works when the CMB spectrum
is slightly larger than the foreground spectrum.
It is not likely that our technique in the current form works
if the foreground spectrum is larger than the CMB one.
However, even in this case, 
it is possible that our technique can help to reduce observational uncertainties
when combined with conventional multi-channel methods.

The spatial filtering that we advocate here may be used as a part of a more 
general filtration procedure which uses both spatial and frequency information. 
Indeed, it is well recognized that the studies of tensor B-modes provide an excessively 
severe challenge to the precision of the removal of foregrounds. 
In this situation, it is important to reduce the errors of the determination of 
the foreground signal. 
The customarily used frequency templates provide filtering which is limited by both 
systematic and measurement errors. 
In this situation, any additional information helping to decrease the errors is highly valuable. 
The spatial power spectrum of the fluctuations that we discussed in the paper does provide 
such an information. 
This approach can be illustrated with Figure~9, which currently has only points for low $l$. 
However, it is already can be seen that the initial error bars of the points can be decreased 
if we require that the points correspond to a power-law spectrum. 
If we 
had more points at higher $l$ it is evident that the uncertainties could be further reduced. 
In other words, the constraint that the foreground fluctuations follow power law allows us 
to partially remove foregrounds at scales at which no foreground templates 
based on multi-frequency measurements are available and, at the same time, 
increase the precision of the foreground removal if the templates are available.

\section{Discussion}

\subsection{Our approach}
The major claim in this paper is that the statistical regularity of the foreground fluctuations enables one
to extend their spectra from the scales where observations are available to the scales with no observations. 
In other words, the spatial spectra of foregrounds are predictable. 
This, in its turn, makes possible spatial filtering of foregrounds. 

The predictability of Galactic foreground fluctuations stems from the fact 
that they are due to ubiquitous Galactic turbulence. 
MHD turbulence is known to have well-defined statistical properties 
both in compressible and incompressible limits (see Cho \& Lazarian 2005 for a review). 
Thus one expects to see power-law behavior, 
which in the case of magnetic field is expected to correspond to 3D power spectrum 
close to the Kolmogorov one. 
In the case of density, the spectrum is Kolmogorov for the subsonic turbulence, 
but gets shallower than the Kolmogorov one for supersonic turbulence (see \S\ref{sect:dust_em}). 
Irrespectively of the underlying 3D spectrum being Kolmogorov or not 
we claim that one can predict the entire spectrum of spatial 2D foreground fluctuations, 
when the measurements over a limited range of scales are available. 
For this purpose we need to know the geometric properties of the volume. 
Our expectations of the underlying spectrum to be in some instances, e.g. for magnetic field, 
to be close to Kolmogorov only help to increase the accuracy of our prediction
of the 2D foreground spectrum. 

With the 2D predicted spectrum one then can use the procedure of filtering 
the foreground using Eq.~(\ref{filter}) (see \S\ref{sect_demo} 
for demonstration of the procedure). 
Alternatively, the good correspondence of the observed spatial spectrum 
of the foregrounds may serve as an additional proof of the accuracy 
of the foreground removal procedure. 
Needless to say, the information of the Galactic turbulence and the geometry of the emitting region 
that can be a by-product of the CMB research is of astrophysical significance. 

A note of warning is due, however. 
The procedure of statistical filtering that we demonstrated in the paper, 
as any other procedure of the foreground removal, is not ideal. 
The power-law approximation of $C_l^F$ is definitely not exact. 
In this paper we have analyzed the causes for such deviations and provided the  
explanation for most notable features characterizing the change. 
More detailed modeling of Galactic turbulence is required. 

\subsection{Our data analysis} 

In this paper, we have discussed angular spectra of Galactic foregrounds.
We have focused on synchrotron total intensity and
polarized thermal dust emission.
Our current study, as well as earlier studies (Chepurnov 1999;
CL02), predicts that $C_l$ will reveal true 3D turbulence spectrum on small
angular scales. 

Our model calculations that take into account stratification effects
imply that
\begin{enumerate}
\item $\theta <$ a few times $0.1^\circ$ (or $l>$ a few times $100$) for synchrotron emission 
(see Fig.~\ref{fig:model}), and
\item $\theta \lesssim 0.1^\circ$ (or $l\gtrsim 1000$) for polarized emission
{}from thermal dust (see Fig.~\ref{fig:starpol}).
\end{enumerate}
On larger angular scales, spectra are expected to be shallower.

\subsection{Results}
In this paper, we have analyzed Haslam 408MHz map and a
model dust emission map and compared the results with model calculations. 
We have found that 
\begin{enumerate}
\item The Haslam map for high Galactic latitude ($b>30^{\circ}$) can be explained by 
      MHD turbulence in the Galactic halo. The measured second-order angular
      structure function is proportional to $\theta^{1.2}$, which corresponds to
      an angular spectrum of $l^{-3.2}$.
      The high-order statistics for high Galactic latitude ($b>30^{\circ}$) is
       consistent with that of incompressible 
       magnetohydrodynamic turbulence.
       Our model calculations show that a two-component model 
      (see \S\ref{sect:3models} and Fig.~\ref{fig:model})
       can naturally explain the observed angular spectrum.
       The one-component model can also explain the observed slope.
       But, the slope of the spectrum shows a more abrupt change
       near $l\sim 30$.

\item The model dust emission map may not have anything to do with turbulence on large angular scales.
      That is, we do not find signatures of turbulence in the map.

\item Both maps show
       flat high-order structure functions for the Galactic plane.
       This kind of behavior is expected
       when discrete structures dominate the map.
\end{enumerate}

We have described how we can obtain angular spectrum of polarized emission from
    thermal dust in high Galactic latitude regions.
Our model calculations show that starlight polarization arising from dust in high Galactic latitude regions
will have a Kolmogorov spectrum, $C_l\propto l^{-11/3}$, 
for $l\gtrsim 1000$ and
a shallower spectrum for $l\lesssim 1000$ (Fig.~\ref{fig:starpol}).
We expect that polarized emission from the same dust also 
has a similar angular spectrum.
That is, we expect that the angular spectrum of polarized emission from thermal dust
is close to a Kolmogorov one for $l\gtrsim 1000$.

We have described a new technique of filtering CMB foregrounds.
When we have
\begin{enumerate}
\item low angular resolution full-sky  measurements (such as WMAP data) and
\item high angular resolution partial-sky measurements with limited frequency channels,
\end{enumerate}
we can use the technique to derive the 
CMB angular spectrum for the high-resolution data.
In \S\ref{sect_demo} and \S\ref{sect_demo2}, we have demonstrated the technique.

\subsection{Comparison with approaches in the literature} \label{sect:6_4}

The existing confusion in the literature include naive identification of the 2D spectra of foregrounds 
with the spectra of the underlying fluctuations. 
Therefore, for instance, from the
fact that the spectral slope of $C_l^F$ differ from the Kolmogorov one, 
the conclusion about the nature of the fluctuations is made. 

In the paper above we have shown that the 2D spectra may have a spectral slope 
different from the underlying spectral slope of the turbulence. 
We showed that it is essential to take into account the non-trivial 
geometry of observations with the observer sampling turbulence along 
the diverging lines of sight and within the volume where the density of emitters changes. 
For the case of the synchrotron fluctuations we showed that the observed non-Kolmogorov value of the spectral index of
$C_l^F$ can be reconciled with the Kolmogorov-type turbulence. 

A notable difference of our study compared with CL02 is that in the latter study 
we tried to necessarily associate the spectra of foregrounds with the underlying Kolmogorov 
or Goldreich-Sridhar (1995) turbulence with the spectral slope $-11/3$. 
The limitations of these approach get evident in view of establishing of the 
shallow spectral index of density fluctuations in supersonic MHD turbulence (Beresnyak et al. 2005). 
These fluctuations, according to our present study, can explain the observed spectra of Galactic dust emission.

\subsection{Limitations and extensions of our filtering procedure}

The procedure of spatial statistical filtering discussed in \S2 is very simplified. 
It should provide satisfactory results for a simple power law behavior of $C_l^F$ 
(see Figure 2). 
In more complex cases detailed modeling of the emitting volume and/or use of the filtering 
as a part of a more sophisticated foreground removal procedure is required. 
We partially addressed this issue in \S\ref{sect_filter}.
Especially in \S\ref{sect_demo2}, we described a method to estimate uncertainties
stemming from a power-law modeling of the foreground spectrum.
Nevertheless, more improvement is needed for the new technique. 
At the same time, we believe that our approach can be applied to the removal of foregrounds 
not only from the CMB data, but, for instance, from the high-$z$ hydrogen statistics studies.

\section{Summary}

In the paper above we have obtained the following results:

1. We provided additional evidence that the synchrotron Galactic emissivity is 
consistent with  the halo + disk model. 
Within this model we show that the spatial spectrum of the underlying 3D 
fluctuations is consistent with the Kolmogorov one.

2. Within our model we related the angular scale for the change of the power spectrum 
of the synchrotron fluctuations with the ratio of the injection scale to the thickness 
of the observed region in the direction of observation.

3. We explained the spectrum of dust foreground emission as arising from the shallow 
spectrum of density fluctuations which characterizes supersonic MHD turbulence.

4. We used numerical  modeling of grain optical properties to relate the polarization 
of starlight with the expected sub-mm foreground polarization and outlined  the ways 
of quantitative use of starlight polarization maps to study sub-mm polarized dust foreground. 
We evaluated the uncertainty of the evaluated sub-mm polarization spectrum arising from the 
variations of the magnetic field direction in respect to the line of sight.

5. We showed that for randomly chosen sample of stars the spectrum of the starlight 
polarization for the underlying Kolmogorov turbulence depended on how the selected stars 
are distributed along the line of sight. For stratified model of Galactic dust 
we predicted the spectrum of spatial fluctuations with the index approaching 
the Kolmogorov value of $-11/3$ for $l>1000$.

6. On the basis of our improved understanding of the self-similar nature of the 
underlying MHD turbulence we proposed and tested a procedure of spatial 
statistical removal of Galactic foregrounds based on extending of the $C_l^F$ spectrum to higher $l$.   

7. We studied the higher order correlations of the Galactic dust and synchrotron 
foregrounds and reported substantial difference with the higher-order scalings (see Appendix D). 
The synchrotron scaling shows intermittencies similar to magnetic fields in incompressible fluids, while dust scaling is closer to the intermittency demonstrated by highly compressible MHD turbulence.

\acknowledgements
J.C.'s work was supported by the National Research Foundation of Korea (NRF)
 grant
funded by the Korean Government (MEST) (NO. 2009-0077372). 
A. Lazarian acknowledges the support by the NSF grants AST 0808118 and AST 0507164, as well as by the
NSF Center for Magnetic Self-Organization in Laboratory and Astrophysical 
Plasmas. 
The works of J.C. was also supported by KICOS through
the grant K20702020016-07E0200-01610 provided by MOST.
We acknowledge the use of the Legacy Archive for Microwave Background Data Analysis (LAMBDA). 
Support for LAMBDA is provided by the NASA Office of Space Science.
We especially thank Dmitri Pogosyan for clarifying the method of estimating the error ranges in
Fig.~\ref{fig:error}(a).
We also thank Alexei Chepurnov, Peter Timbie, John Everett and Simon Prunet for valuable discussions.

\appendix
\section{Spectrum and Structure Function of Diffuse Synchrotron Emission: 
a model for homogeneous turbulence revisited} \label{sect:spsfco}
Suppose that 3-dimensional (3D) MHD turbulence
has a 3D spatial power spectrum of the form $E_{3D}\propto k^{-m}$, 
where $k$ is the wavenumber. Note that in Kolmogorov turbulence
$m=11/3$.
Then what will be the 
2-dimensional angular spectrum, $C_{l}$, 
of the observed synchrotron total intensity?
We cannot not directly observe the 3D spatial power spectrum.
However, we can infer the 3D spectrum from the observed
2-dimensional (2D) angular spectrum, $C_{l}$, of
the synchrotron total intensity.
Then let us find out how the 3D spectrum, $E_{3D}(k)$,
and the 2D angular spectrum, $C_l$, are related.
We mostly follow discussions in Cho \& Lazarian (CL02).
We also make use of an analytical insight
obtained in Lazarian (1992, 1995ab) and numerical results
obtained in CL02.
Although we focus on synchrotron emission here, 
the discussion in this section
can be applicable to any kind of emission
{}from an optically thin medium.

\subsection{MHD turbulence and synchrotron emission} \label{sect:synchem}

For synchrotron radiation, emissivity at a point ${\bf r}$ is given by
   $\epsilon({\bf r}) \propto n(e) |{B}_{\bot}|^{\gamma}$,
where $n(e)$ is the electron number density, $B_{\bot}$ is the component
of magnetic field perpendicular to the line of sight. The index
$\gamma$ is approximately $2$ for radio synchrotron frequencies
(see Smoot 1999).
If electrons
are uniformly distributed over the scales of magnetic field inhomogeneities,
the spectrum of synchrotron
intensity reflects the statistics of magnetic field.
For small amplitude perturbations
($\delta b/B\ll 1$; this is true for scales several times 
smaller than the outer scale of turbulence 
if we interpret $B$ as local mean magnetic field strength and $\delta b$ as random
fluctuating field in the local region), 
if $\delta b$
has a power-law behavior, the synchrotron emissivity will have the 
same power-law behavior (see Getmantsev 1959; Lazarian \& Shutenkov 1990;
Chepurnov 1999).
Therefore, we expect that the angular spectrum of synchrotron
intensity also reflects the spectrum of 3-dimensional MHD turbulence.

When an observer is located inside a turbulent medium, the angular 
correlation function, hence the power spectrum,
 shows two asymptotic behaviors. When the angle is larger than
a critical angle, we can show that 
the angular correlation shows a universal $\theta^{-1}$
scaling. On the contrary, when the angular separation
 is smaller than the critical angle, 
the angular correlation
reflects statistics of turbulence. In this small angle limit,
we can show that
the angular power spectrum is very similar to that of turbulence.
The critical angle is determined by the geometry.
Let the outer scale of turbulence be $L$ and the distance to the farthest
eddies be $d_{max}$.
Then the critical angle is 
\be
  \theta \sim L/d_{max}.
\ee

\subsection{Small-angle limit in homogeneous turbulence} \label{sect:smangle}
When the angle between the lines of sight
is small (i.e.~$\theta < L/d_{max}$), the
angular spectrum $C_l$ has the same slope as the 3-dimensional
energy spectrum of turbulence.
 Lazarian \& Shutenkov (1990) showed that if the 3D spatial spectrum
of a variable follows
a power law, $E_{3D}(k)\propto k^{-m}$, 
then the 2-dimensional spectrum of the variable projected on the sky also
 follows the same power law,
\begin{equation}
   C_l \propto   l^{-m}
\end{equation}
in the small $\theta$ limit.
For Kolmogorov turbulence ($E_{3D}\propto k^{-11/3}$),
we expect
\begin{equation}
C_l \propto l^{-11/3}, \mbox{~~~if $\theta<L/d_{max}$.}
 \label{eq_5}
\end{equation}
Note that $l\sim \pi/\theta$.

In some cases, when we have data with incomplete sky coverage,
we need to infer $C_l$ from the observation of 
    the angular correlation function 
\be
  w(\theta)=< I({\bf e}_1) I({\bf e}_2) >,
\ee
where $I({\bf e})$ is the intensity of synchrotron emission,
${\bf e}_1$ and ${\bf e}_2$ are unit vectors along the lines of
sight, $\theta$ is the angle between ${\bf e}_1$ and ${\bf e}_2$, and
the angle brackets denote average taken over the observed region.
As we discuss in Appendix A, when the underlying 3D turbulence spectrum
is $\propto k^{-m}$ (e.g. $m=11/3$ for Kolmogorov turbulence),
the angular correlation function $w(\theta)$ is given by
\be
   w(\theta) ~\propto ~< I^2 >-\mbox{const}~\theta^{m-2}, \mbox{~~~if $\theta < L/d_{max}$.}
\ee
It is sometimes
inconvenient to use the angular correlation function in practice to
study turbulence statistics
because of the constant $< I^2 >$.

A better quantity in small-angle
limit would be the second-order angular structure function:
\bea
   D_2 (\theta) & = & < | I({\bf e}_1) -I({\bf e}_2) |^2 > \\
     & = & 2< I^2 > - 2w(\theta).
\eea
Thus, in homogeneous turbulence with 3D spatial spectrum 
of $E(k)\propto k^{-m}$, we have
\be
   D_2 (\theta) \propto \theta^{m-2}.
\ee
When we measure the slope of the angular structure function, we
can infer the slope of the 3D spatial power spectrum of turbulence.

\begin{figure}[h!t]
\includegraphics[width=0.40\textwidth]{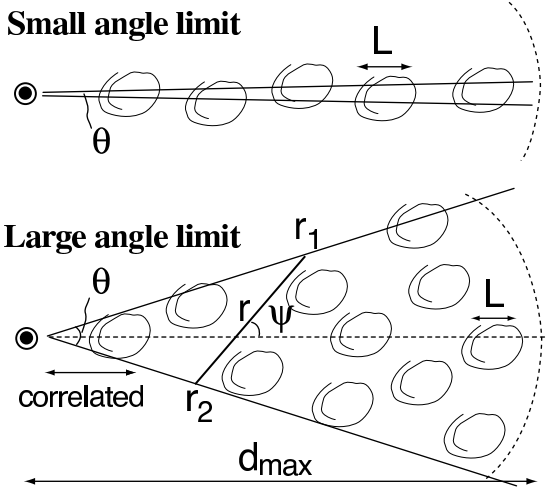}  
\caption{Two limits in homogeneous turbulence.
  {\it Upper plot}:
Small $\theta$ limit ($\theta < L/d_{max}$). 
The fluctuations along the entire length of the lines of sight
are correlated. 
{\it Lower plot}: Large $\theta$ limit ($\theta > L/d_{max}$). 
Only points close to the observer
are correlated. Note the definition of
$r$ and $\psi$. From CL02.
}
\label{fig:0}
\end{figure}

\subsection{Large-angle limit in homogeneous turbulence}
In this limit, the angular correlation function is 
more useful than the structure function.
{}Following Lazarian \& Shutenkov (1990), we can show that
the angular correlation function for $ \theta > L/d_{max}$ follows
\begin{eqnarray}
    w(\theta) & = & \int \int dr_1 dr_2 ~{\cal K}( |{\bf r}_1-{\bf r}_2| ), \nonumber
\\
              & = & \frac{1}{\sin{\theta}}
                    \int_0^{\infty}dr ~r {\cal K}(r) \int_{\theta/2}^{\pi-\theta/2}d\psi 
\propto \frac{\pi-\theta}{\sin\theta} \sim \frac{const}{\theta},
\end{eqnarray}
where ${\cal K}(r)$ is the 3D spatial correlation function and
we change variables: $(r_1,r_2)\rightarrow (r,\psi)$, 
which is clear from
Fig.~\ref{fig:0}. We accounted for the Jacobian of which is $r/\sin{\theta}$.
We can understand $1/\theta$ behavior qualitatively as follows.
When the angle is large, points along  of the lines-of-sight near the observer
are still correlated. These points extend from the observer
over the distance $\propto 1/\sin{(\theta/2)}$.

If we assume $L/d_{max}< \theta\ll 1$,
we can get the angular power spectrum $C_l$ 
using Fourier transform:
\begin{eqnarray}
   C_l & \sim &  \int \int  w(\theta) 
                 e^{-i{\bf l}\cdot {\bf \theta}} d\theta_x d\theta_y  
\nonumber \\
       & \sim &  \int d\theta ~\theta J_0(l\theta) w(\theta) \propto   l^{-1},
   \label{eq_7}
\end{eqnarray}
where $\theta=(\theta_x^2+\theta_y^2)^{1/2}$, $J_0$ is the Bessel function, and
we use $w(\theta)\propto \theta^{-1}$.

\subsection{Expectations for homogeneous turbulence}
In summary, for homogeneous Kolmogorov turbulence (i.e.~$E(k)\propto k^{-11/3}$), 
we expect from equations (\ref{eq_5}) and
(\ref{eq_7}) that
\begin{equation}
 C_l \propto \left\{ \begin{array}{ll} 
                         l^{-11/3}     & \mbox{if $l>l_{cr}$} \\
                         l^{-1}        & \mbox{if $l<l_{cr}$,}
                      \end{array}
              \right.
     \label{eq_1_11_3}
\end{equation}
which means that the power index $\alpha$ of $C_l$ is\footnote{ 
      Note that point sources would result in $\alpha \sim 0$.} 
$-1 \leq \alpha \leq -11/3$.
We expect the following scaling for 
the second-order angular structure function:
\begin{equation}
 D_2(\theta) \propto \left\{ \begin{array}{ll} 
                           \theta^{5/3}   & \mbox{if $\theta <L/d_{max}$ } \\
                         \mbox{constant}  & \mbox{if $\theta >L/d_{max}$.}
                      \end{array}
              \right.
     \label{eq_d2}
\end{equation}
The critical angle $\theta_{cr} \sim L/d_{max}$ 
depends on the size of the large
turbulent eddies and on the length of the line of sight. 
If we assume that turbulence is homogeneous along the lines
of sight and has $L\sim 100\ pc$ corresponding to a typical size of the
supernova remnant, and that $d_{max}\sim 1$~kpc for synchrotron halo (see Smoot 1999), 
we get $\theta_{cr} \sim 6^{\circ}$.

\section{The  second-order angular structure function in the small angle limit}
In this Appendix, we discuss how underlying 3D statistics and observed 2D statistics are related.
This discussion is useful when we infer 3D statistics from observed 2D statistics, or vice versa.
Strictly speaking, discussion in the section is applicable to homogeneous 
and isotropic turbulence only.\footnote{
    When turbulence is inhomogeneous or anisotropic, we may not directly apply
    the results in this section. However, our numerical calculations in \S\ref{sect:3models}
    shows that the relation between 3D statistics and 2D angular correlation function (or
    2D angular structure function) discussed in this section is also applicable to
    inhomogeneous cases.}

The angular correlation $w(\theta)$ is given by the integral
\be
   w(\theta)=\int dl_1 \int dl_2 ~{\cal K}(|{\bf l}_1 - {\bf l}_2|),
\ee
where ${\cal K}(r)$ is the 3 dimensional spatial correlation.
Suppose that ${\bf l}_1$ is along x-axis, ${\bf l}_1=(l_1,0)$,  and 
${\bf l}_2 = l_2 (\cos \theta, \sin \theta)$.
Then, the correlation in the limit of small $\theta$ is given by
\bea
   w(\theta) & = &\int dl_1 \int dl_2 
              ~{\cal K}\left( \sqrt{ (l_1-l_2 \cos\theta)^2 + l_2^2 \sin^2\theta } 
      \right) \\
      & = & \int dl_1 \int dl_2 
              ~{\cal K} \left( \sqrt{ (l_1^2-2l_1 l_2 \cos\theta + l_2^2 } \right) \\
      & \approx & \int dl_1 \int dl_2
              ~{\cal K} \left( \sqrt{ (l_1- l_2)^2 + l_1 l_2\theta^2 } \right).
\eea
Suppose that the spatial correlation follows a power law: ${\cal K}(r) \propto \mbox{~const}-r^p$ 
for $r < L$, where $L$ is the outer-scale of turbulence.
For Kolmogorov turbulence, $p=2/3$.
Then the derivative of $w(\theta)$ is given by
\bea
  \frac{ w(\theta) }{ d\theta } & \propto &
      -\int dl_1 \int dl_2
              ~\left[  (l_1- l_2)^2 + l_1 l_2\theta^2  \right]^{p/2-1} (2l_1l_2\theta) \\
   & -\propto &
      \int du \int dw
              ~\left[  w^2 + (u^2-w^2) \theta^2 /4 \right]^{p/2-1} (u^2-w^2)\theta /2 ,
\eea
where $u=l_1+l_2$ and $w=l_1-l_2$.
If $p \leq 1$, the integration diverges as $\theta$ goes to zero.\footnote{
    When $p=1$, the spatial correlation becomes ${\cal K}(r)\propto C-r$, where
    $C$ is a constant.
    The corresponding 3D spectrum is $E(k) \propto k^{-4}$.
    When the slope of the turbulence spectrum is 
    steeper than $k^{-4}$,
    the correlation function has the form 
    ${\cal K}(r) \propto {\cal K}_0-r^{1}$ regardless
    of the turbulence slope.
    On the other hand, when the three-dimensional spectrum of turbulence 
    is shallower than $k^{-4}$, we have
    ${\cal K}(r) \propto {\cal K}_0-r^{m-3}$, where ${\cal K}_0\sim L^{m-3}$
    is a constant.
    Therefore, the condition of $p\leq 1$ is generally satisfied in turbulent
    medium.
}
Therefore, when $p \leq 1$, it suffices to perform the integration in the vicinity
of $l_1=l_2$ or $w=0$. Then we have
\bea
 \frac{ w(\theta) }{ d\theta } & \propto &
     -\int du \int dw
             ~\left[  w^2 + u^2 \theta^2 /4 \right]^{p/2-1} u^2\theta /2 \\
    & \approx & -\int du ~(u \theta/2)^{p-1} u^2 \theta /2
     \propto  -\theta^p,
\eea
where we use $\int_{-\infty}^{+\infty} dw/(w^2+A^2)^n = 
    A^{1-2n}\int_{-\pi/2}^{+\pi/2} d\theta ~\sec^{2-2n}\theta $.
Therefore, for small $\theta$ we have
\be
  w(\theta) \propto C_1 - C_2\theta^{p+1},
\ee
where $C_1$ and $C_2$ are constants.
Comparing this equation with 
\be
  w(\theta)=\mbox{$C_3$}-C_4 D_2(\theta),
\ee
we get
\be
   D_2(\theta) \propto \theta^{p+1}.
\ee

Analytic expressions for the relation between the angular structure
function ($D_2$) and the spatial 1D spectrum (E(k); in case of Kolmogorov, 
$E(k)\propto k^{-5/3}$) 
can be found in the literature.
For example, Lazarian (1995a; see also Lazarian \& Shutenkov 1990) 
derived the following expression:
\be
 E(k) \propto k \int_0^{{\cal L}/R} d\eta~ \frac{ d }{ d\eta } 
    \left( Q(\eta)\eta \right) J_1(kR\eta) + {\cal K}_5,
\ee
where ${\cal L}$ can be regarded as the outer scale of turbulence,
$R$ is the size of the system,
$Q(\eta)\sim D_2^\prime(\eta)\eta$, $\eta=\sin\theta$, $J_1(x)$
is the Bessel function of the first order, and ${\cal K}_5$
is a small correction term.

\section{Spatial spectrum of emissivity}
The synchrotron emissivity is proportional to $\sim n(e)B^\gamma \propto B^2$, where
$n(e)$ is the high-energy electron number density.
Suppose that magnetic field is roughly a Gaussian random variable.
This may not be exactly true, but should be a good approximation.
When a Gaussian random variable $B({\bf r})$\footnote{
    For simplicity, we assume $B$ is a scalar.
}
follows a Kolmogorov spectrum
\be
  E_{B,3D}\equiv |\tilde{B}(k)|^2 \propto \left\{ \begin{array}{ll} 
                              0                        & \mbox{if $k\le k_0$} \\
                              (k/k_0)^{-11/3}          & \mbox{if $k\ge k_0$,}
                      \end{array}
              \right. \label{B_E_3D}
\ee
we can show that the 3D spectrum of $B^2(r)$ follows Eq.~\ref{E_3D} (see, for example,
Chepurnov 1999).
The correlation of $B^2(r)$ and 3D energy spectrum of $B^2(r)$ are related by
\bea
     {\cal K}_{B^2}({\bf r})=< B^2({\bf x})B^2({\bf x}+{\bf r}) >_x \propto 
          \int  E_{B^2, 3D}({\bf k}) e^{i{\bf k}\cdot {\bf r}} d^3{\bf k}, \\
     E_{B^2, 3D}({\bf k})\equiv |\tilde{B^2}(k)|^2 \propto 
           \int {\cal K}_{B^2}({\bf r}) 
               e^{-i{\bf k}\cdot {\bf r}} d^3{\bf r},  \label{eq:cor2sp}
\eea
where $<...>_x$ denotes an average over ${\bf x}$.
A Gaussian random variable satisfies
\be
        < B^2({\bf x})B^2({\bf x}+{\bf r}) >
     = < B^2({\bf x})><B^2({\bf x}+{\bf r}) > 
      +2<B({\bf x}) B({\bf x}+{\bf r})>^2,
\ee
where the first term on the right is a constant.
Therefore we can ignore the term in what follows.
Fourier transform of both sides results in
\bea
   \mbox{LHS} & = & E_{B^2, 3D}({\bf k}), \\
   \mbox{RHS} & = & 2 \int <B({\bf x})B({\bf x}+{\bf r})>^2 
                  e^{-i{\bf k}\cdot {\bf r}} d^3{\bf r} \\
          & = & 2\int~d^3{\bf r} ~{\cal K}_{B}({\bf r})~{\cal K}_{B}({\bf r}) 
                  e^{-i{\bf k}\cdot {\bf r}}  \\
          & = & 2\int~d^3{\bf r} \int~d^3{\bf p} \int~d^3{\bf q}
                  ~E_{B, 3D}({\bf p})E_{B, 3D}({\bf q})
                   e^{i({\bf p}+{\bf q}-{\bf k})\cdot {\bf r}}  \\
          & = & 2 \int~d^3{\bf p} \int~d^3{\bf q}
                  ~E_{B, 3D}({\bf p})E_{B, 3D}({\bf q})
                   \delta( {\bf p}+{\bf q}-{\bf k})  \\
          & = & 2\int~d^3{\bf p} ~E_{B, 3D}({\bf p})~E_{B, 3D}({\bf k}-{\bf p}),
\eea
where $\delta(k)$ is the Dirac $\delta$-function.
Therefore we have
\be
    E_{B^2, 3D}({\bf k}) \approx E_{B^2, 3D}({\bf 0})
    \approx 2\int~d^3{\bf k} | E_{B, 3D}({\bf k}) |^2
    \approx \mbox{constant}
\ee
for $k \ll k_0$.

\section{High-order statistics}  \label{sect:h_o}

While most of the paper is directly related to making use of the knowledge
of the underlying spectra and/or two points correlations in order to remove
foregrounds, the part dealing with higher-order statistics is not {\it directly}
related to the foreground removal. Nevertheless, the correspondence of the intermittencies of 
foregrounds to those of turbulence provide another support for our understanding 
of the turbulent origin of foreground fluctuations. 

High-order structure functions 
are used for the study of intermittency, which refers to the non-uniform distribution of structures.
Since CMB signals are close to Gaussian, one may think that they do not
 have strong intermittency.
However, there is a report that intermittency of CMB signals
deviate from Gaussianity (Bershadskii \& Screenivasan 2003).
If intermittency of foreground signals are different from that of CMB signals,
we can potentially use high-order structure functions to separate
CMB and foreground signals.
In addition, we can use high-order structure functions to improve our knowledge of
foregrounds.

The structure functions of order $p$ for an observable $I$ is defined by
\be
   S_p(r)=< |I(x)-I(x+r)|^p>,
\ee
where the angled brackets denote average over position $x$.
For an observable defined in the plane of the sky, 
the angular structure function of order $p$ is
\be
   D_p(\theta)=< |I({\bf e}_1)-I({\bf e}_2)|^p>.   
\ee

Traditionally, researchers use
high-order structure functions of velocity to probe 
dissipation structures of turbulence.
In fully developed hydrodynamic turbulence, the (longitudinal)
velocity structure functions
$S_p=< ( [ {\bf v}({\bf x}+ {\bf r}) -
      {\bf v}({\bf x})]\cdot \hat{\bf r} )^p>
\equiv < \delta v_L^p({\bf r}) >$ are
expected to scale as $r^{\zeta_p}$:
\be
  S_p(r)\propto r^{\zeta_p}.
\ee
One of the key issues in this field is the functional form of the
scaling exponents $\zeta_p$.
There are several models for $\zeta_p$.
Roughly speaking, the dimensionality 
of the energy dissipation structures plays an important role.

Assuming 1-dimensional worm-like 
dissipation structures, She \& Leveque (1994) 
proposed a scaling relation
\be
  \zeta_p^{SL}=p/9+2[1-(2/3)^{p/3}]
\ee
for incompressible hydrodynamic turbulence.
Note that $\zeta_p$ is the scaling exponent of structure of order $p$.
On the other hand, assuming 2-dimensional sheet-like dissipation structures,
M\"uller \& Biskamp (2000) proposed the relation
\begin{equation}
\zeta_p^{MB}=p/9+1-(1/3)^{p/3}
\end{equation}
for incompressible magneto-hydrodynamic turbulence\footnote{
   Boldyrev (2002) obtained the same scaling relation for 
   highly supersonic turbulence. However, since it is unlikely that
   turbulence in the Galactic halo is highly supersonic, we refer the
   scaling relation to 
   the ``Muller-Biskamp'' scaling.
}.

Recently, high-order structure functions of molecular line intensities
have been also employed (Padoan et al. 2003; 
Gustafsson et al. 2006).
In optically thin case, the molecular line intensities are proportional
to the column density.
Kowal, Lazarian, \& Beresnyak (2007) studied scaling of higher moments of
density fluctuations in MHD turbulence.
Their numerical results show that, first of all, the scalings of higher moments in 3D can be obtained by studying 
the distribution of column densities. Then, they showed that the behavior of the scaling exponents 
for column density depends on sonic Mach number of turbulence.

Bershadskii \& Screenivasan (2003) calculated intermittency of
CMB signals. Their result shows that the WMAP data follow
She-Leveque scaling. Therefore, it is interesting to see
if foreground signals show different scalings.

\begin{figure*}[h!t]
\includegraphics[width=0.32\textwidth]{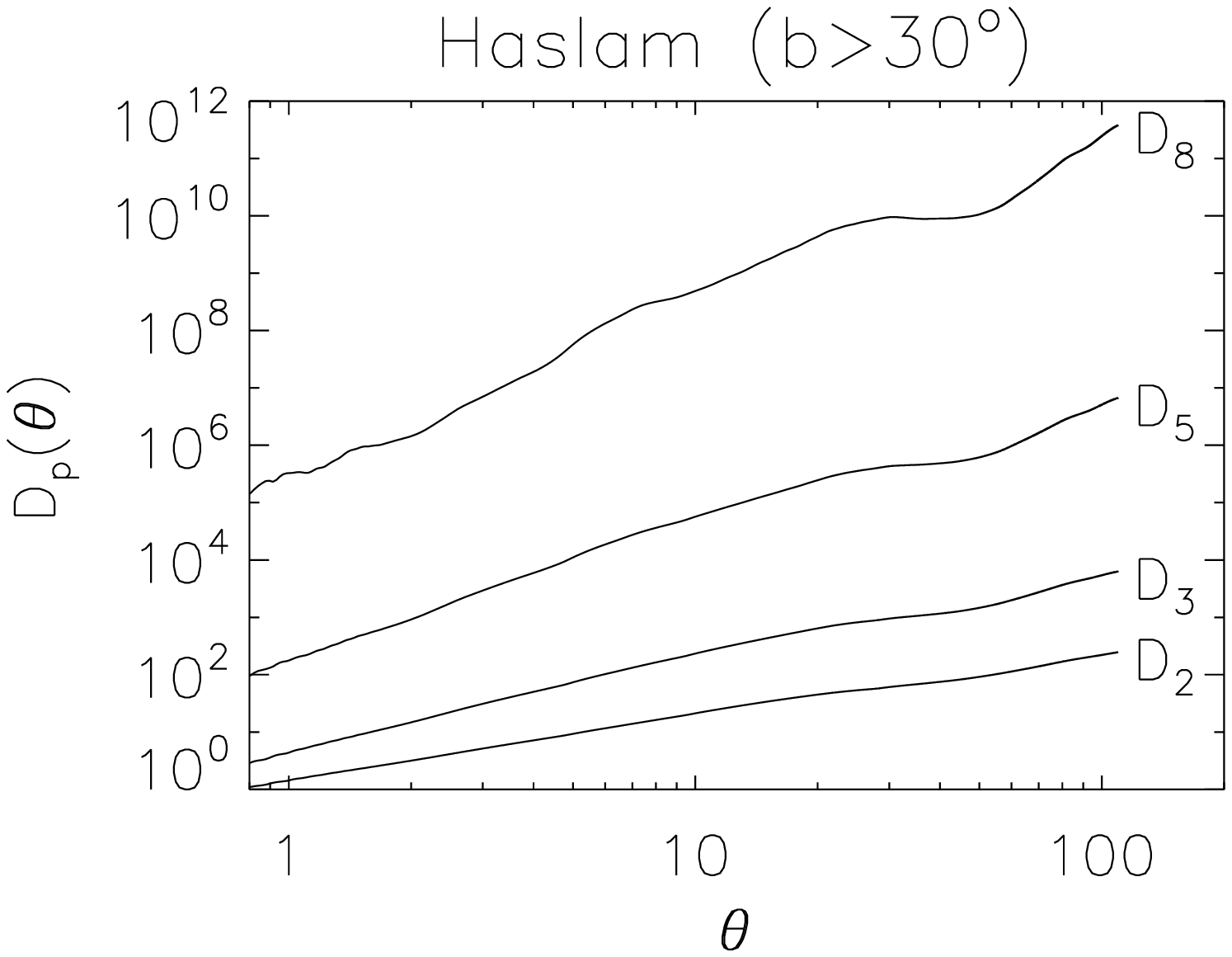}   
\includegraphics[width=0.32\textwidth]{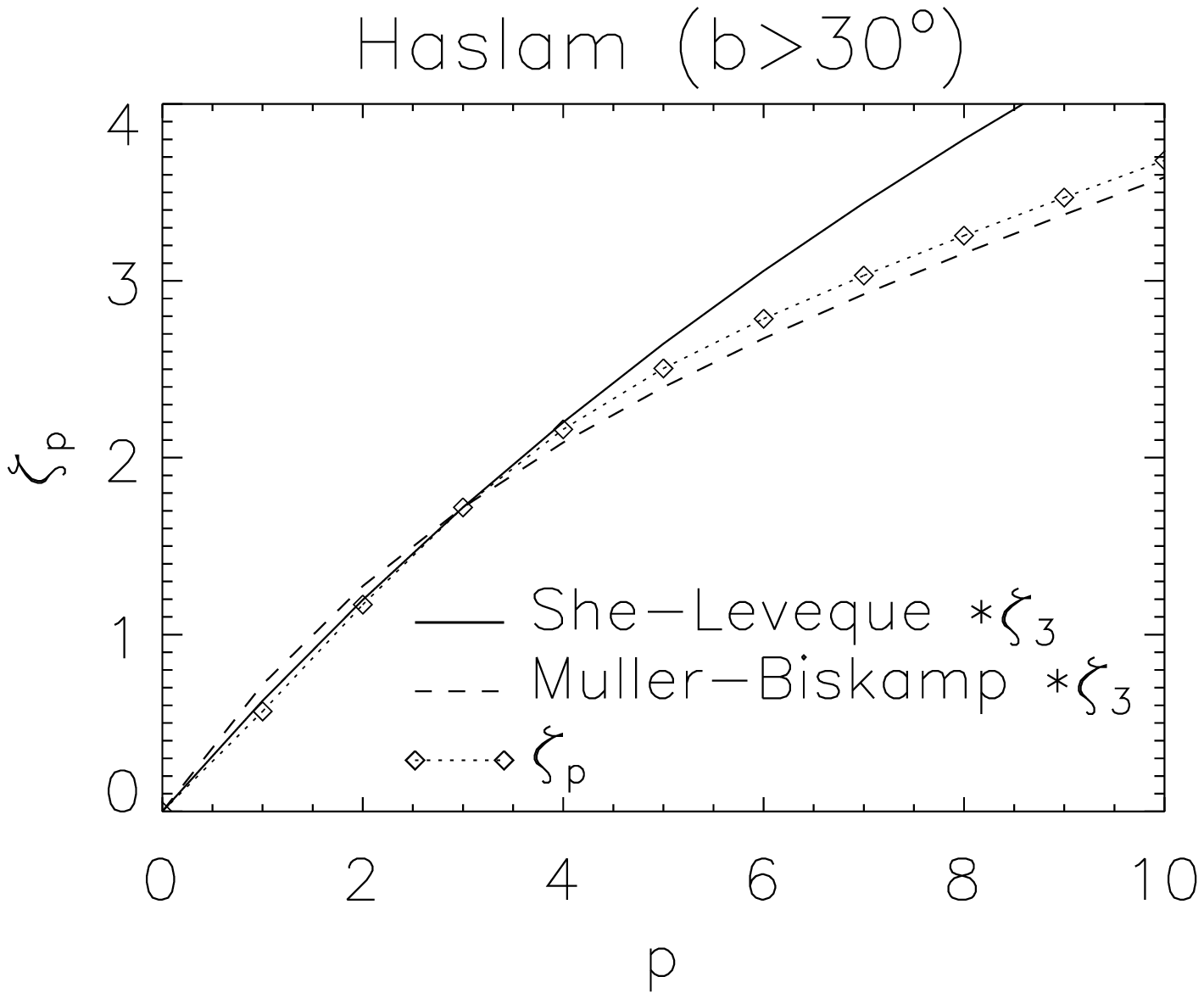}  
\includegraphics[width=0.32\textwidth]{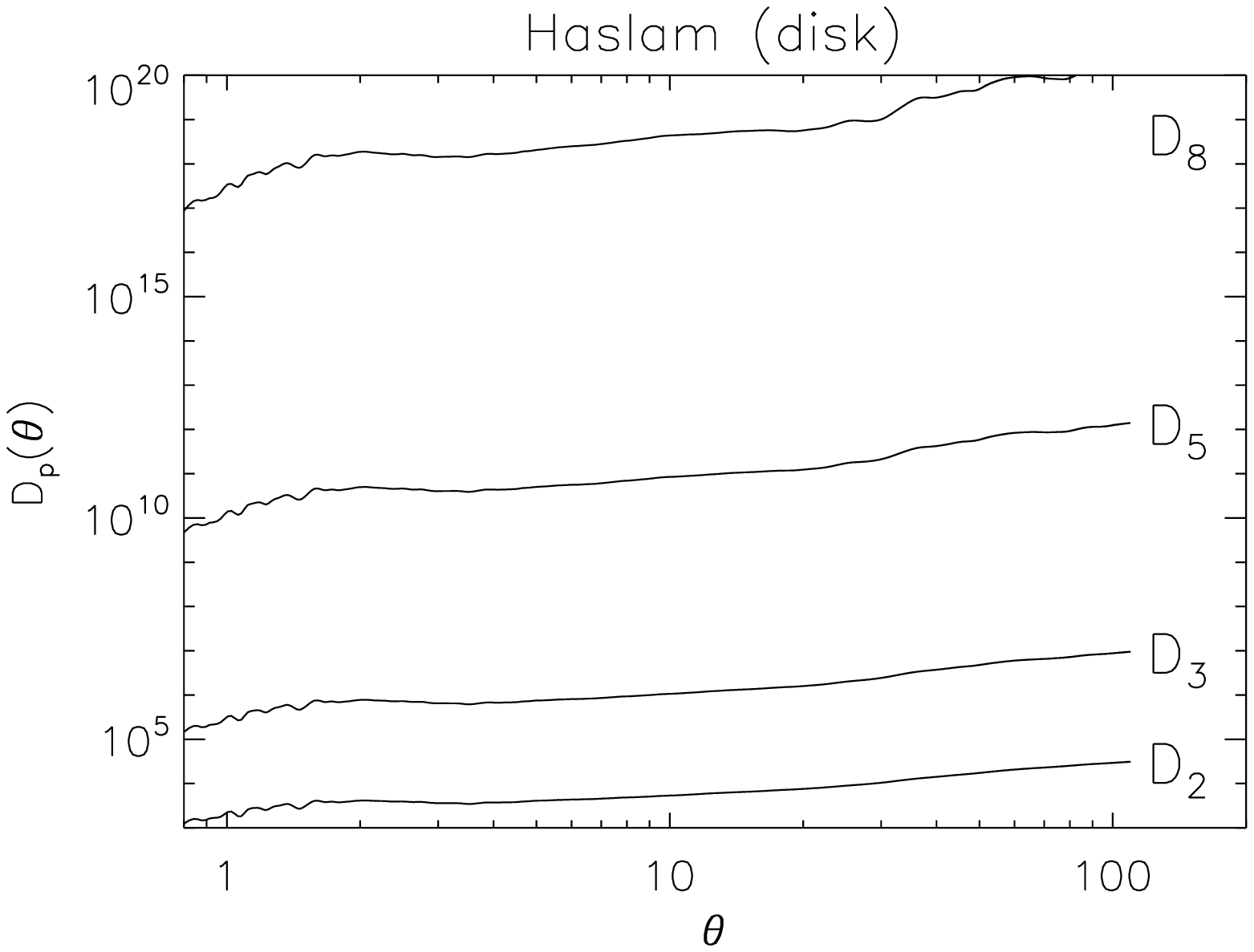}   
\caption{ 
   High order statistics of the Haslam 408MHz map.
 {\it Left}: The second, third, 5th, and 8th-order structure functions
     for $b>30^{\circ}$.
 {\it Middle}: The scaling exponents for $b>30^{\circ}$
      seem to follow the Muller-Biskamp  
      MHD scaling.
      Note that we equate the scaling exponent of the observed third-order
      structure function and that of the Muller-Biskamp model (see more discussion in 
Cho et al. 2003).
      We measure the slope between $\theta=1^{\circ}$ and $10^{\circ}$.
 {\it Right}: The second, third, 5th, and 8th-order structure functions
     for the Galactic disk ($ -2^\circ \le b \le 2\circ$).
     They are all flat.
}
\label{fig_10}
\end{figure*}

We plot the scaling exponents of the Haslam 408MHz map in Fig.~\ref{fig_10}.
The Haslam map shows a reasonable agreement with the Muller-Biskamp 
MHD model.
  Note that Padoan et al. (2003) also obtained 
a similar result using $^{13}CO$ emission from Perseus and Taurus.

Unlike the Haslam map, the 94GHz dust map does not show agreement with
Muller-Biskamp model. The scaling exponents do not show strong dependence
on the order $p$ similar, which makes them similar to the results of the 
potential part of the supersonic flow studied in Kowal \& Lazarian (2010) and
densities in supersonic super-Alfvenic flows studied in Kowal et al. (2007). One
may expect that in the presence of self-gravity this is what is expected for
higher order statistics of density, but further studies are required.  
 
\begin{figure*}[h!t]
\includegraphics[width=0.32\textwidth]{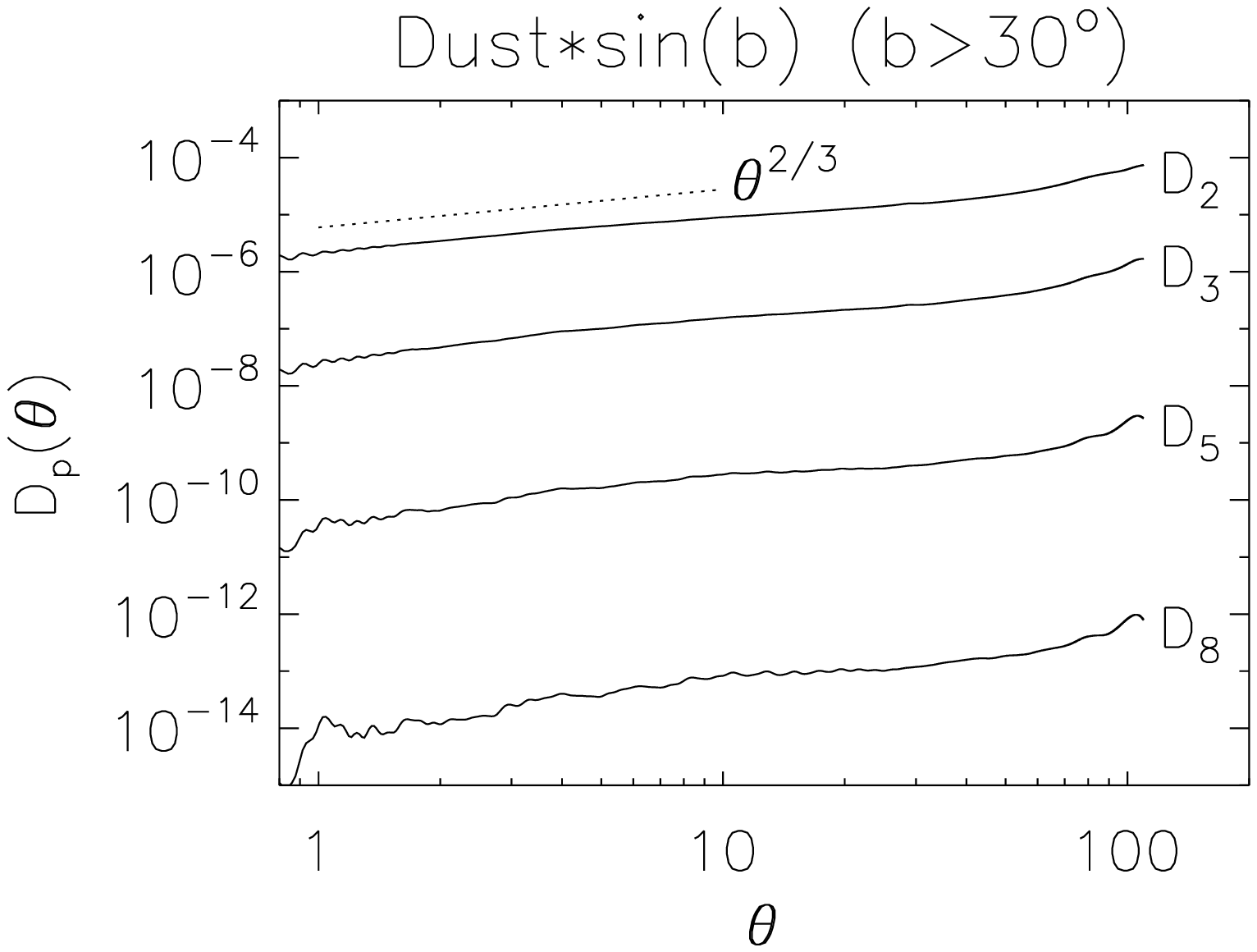}   
\includegraphics[width=0.32\textwidth]{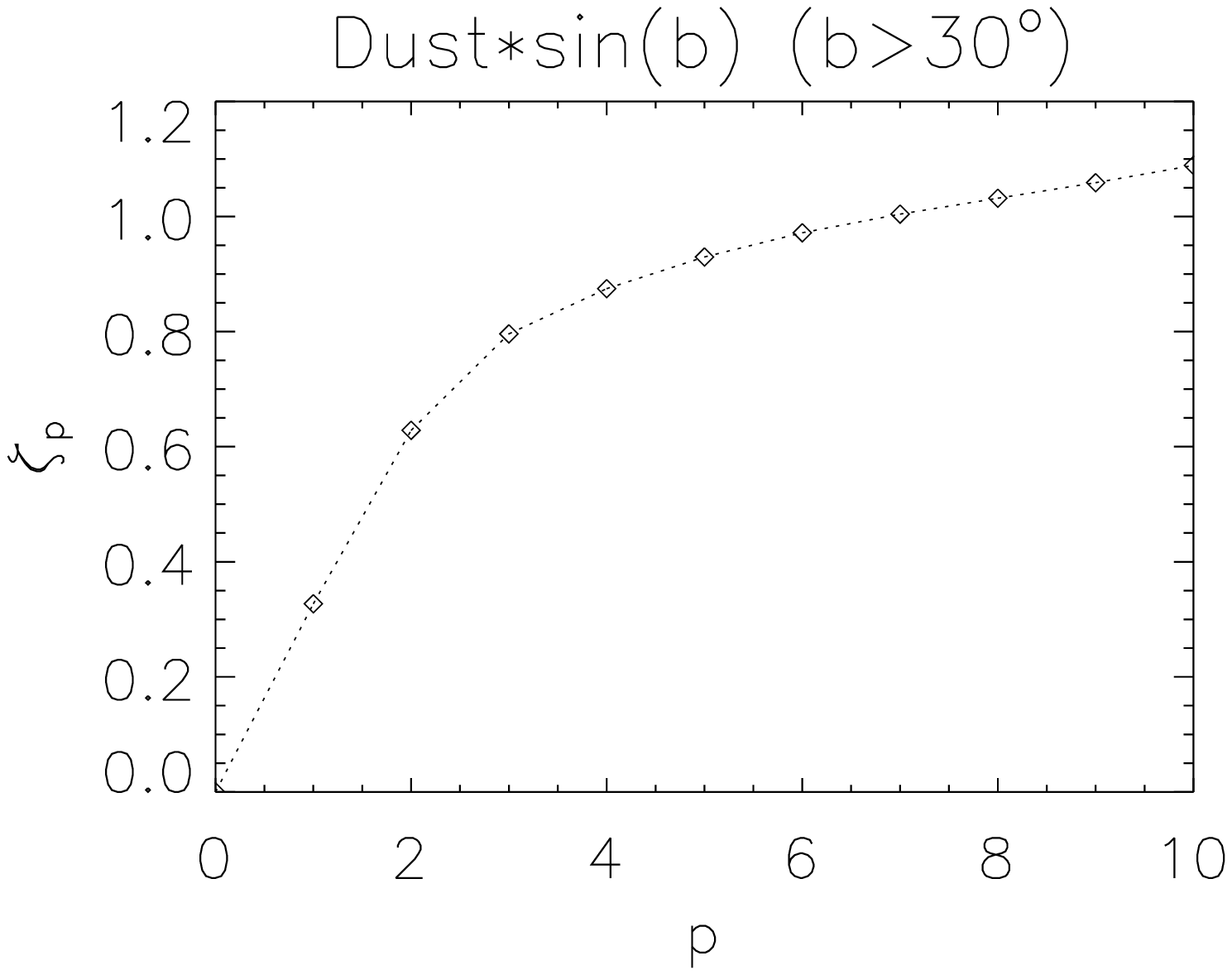}   
\includegraphics[width=0.32\textwidth]{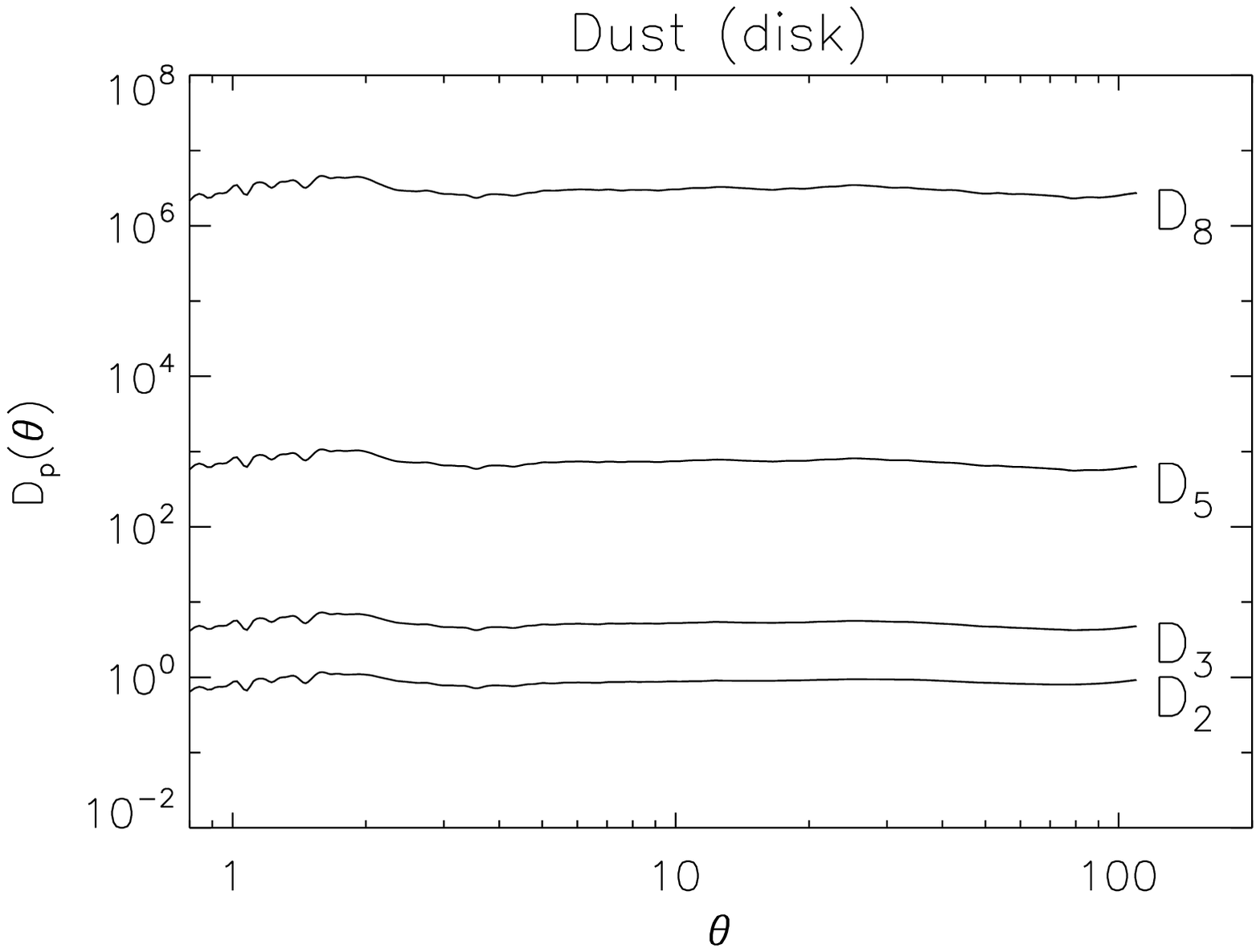}   
\caption{ 
   High order statistics of the 94GHz dust emission map. 
   We use dust emission intensity times $\sin b$ for all calculations.
 {\it Left}: The second, third, 5th, and 8th-order structure functions
     for $b>30^{\circ}$. 
 {\it Middle}: The scaling exponents for $b>30^{\circ}$
      does not show signatures of turbulence.
      We measure the slope between $\theta=1^{\circ}$ and $10^{\circ}$.
 {\it Right}: The second, third, 5th, and 8th-order structure functions
     for the Galactic disk ($ -2^\circ \le b \le 2\circ$).
     They are all flat.
}
\label{fig_11}
\end{figure*}

Left and middle panels of Fig.~\ref{fig_11} 
   show that the slope is around 1 for high-order structure functions.
This kind of behavior is expected
when discrete structures dominate the map.
However, it is not clear what kinds of discrete structures dominate\footnote{
   However, we can show that thin filamentary structures or point sources 
  {\it with
   uniform intensity and sharp boundary} are not the dominant structures.
   Consider a circular cloud with
   a radius $\Delta$ and a uniform intensity $I$ centered
   at the origin.
   (For simplicity, let us consider a two-dimensional cloud 
    in the two-dimensional Cartesian coordinate system.
    Let us assume intensity is defined on the Cartesian grid.)
   Then the structure function is given by
   $ D_n(r) \propto \sum_i \sum_j |I({\bf x}_i)-I({\bf x}_j)|^n /N_{pair},$
   where two points ${\bf x}_i$ and ${\bf x}_j$ are separated by the distance $r$
   and $N_{pair}$ is the total number of such pairs.
   When $r\gg \Delta$, 
   we can show that $\sum_i \sum_j |I({\bf x}_i)-I({\bf x}_j)|^n \propto 
   I^n (2\pi r)(\pi \Delta^2)$ and $N_{pair} \propto (2\pi r)$.
   Therefore, we have $D_n(r) \propto I^n (\pi \Delta^2)$
   for $r\gg \Delta$. That is, the structure functions show no dependence on $r$.
   We expect that structure functions for a thin uniform filament also show 
   no dependence on $r$ when $r$ is larger than the width of the filament,
   because a filament can be viewed as a chain of
   circular clouds (or a chain of square-like clouds).
   The high-order structure functions for the dust emission map
   show $\sim r^1$ power-law scaling
   for $\theta \gtrsim 1^\circ$ (left panel of Fig.~\ref{fig_10}).
   Thus  neither thin uniform filaments
   nor point sources are dominant structures.
   However, filaments or circular clouds with smoothly extended boundaries
   can be the dominant structures.
}.
Since the Haslam map and the dust map sample different types of 
the ISM, it is not
so surprising that they show different scaling behaviors.

For the Galactic disk, high-order structure functions
of both maps show nearly flat structure functions.
The structure functions show a nearly flat behavior 
down to $\theta \sim 1^\circ$, which happens to be the actual
angular resolution of the maps. Further studies of MHD turbulence
in the presence of self-gravity should clarify the origin of such behavior.
Alternatively, higher moments of structure functions  can be affected by 
unresolved point sources. 


\end{document}